\documentclass{article}

 \usepackage[main, final]{neurips_2026}

\usepackage[utf8]{inputenc} 
\usepackage[T1]{fontenc}    
\usepackage{hyperref}       
\usepackage{url}            
\usepackage{booktabs}       
\usepackage{amsfonts}       
\usepackage{amsmath}
\usepackage{amssymb}     
\usepackage{mathtools}   
\usepackage{physics}
\usepackage{nicefrac}       
\usepackage{microtype}      
\usepackage{xcolor}         
\usepackage{booktabs}      
\usepackage{algorithm}
\usepackage{algpseudocode}
\usepackage{amsmath}
\usepackage{multirow}      
\usepackage{graphicx}      
\usepackage{colortbl}      
\usepackage{pifont}        
\usepackage{amssymb}       
\usepackage{makecell}
\usepackage{listings}
\usepackage{inconsolata}
\definecolor{boxgreen}{HTML}{00645F}
\definecolor{boxbg}{HTML}{F1F8F8}
\definecolor{tokenred}{HTML}{FF1A1A}
\definecolor{tokenorange}{HTML}{FF7F00}
\definecolor{tokenblue}{HTML}{0000FF}
\usepackage{listings}
\usepackage{tcolorbox}
\tcbuselibrary{listings}
\usepackage{subcaption}
\definecolor{boxgreen}{HTML}{00645F}
\definecolor{boxbg}{HTML}{F1F8F8}

\newtcolorbox{promptbox}[1][]{
  colback=gray!5,
  colframe=gray!60,
  boxrule=0.6pt,
  arc=2mm,
  left=2mm,
  right=2mm,
  top=1mm,
  bottom=1mm,
  fonttitle=\bfseries,
  title=#1
}

\newcommand{\cmark}{\textcolor{green!60!black}{\ding{51}}}   
\newcommand{\xmark}{\textcolor{red!70!black}{\ding{55}}}     
\newcommand{\titleicon}{
    \raisebox{-0.04\height}{\includegraphics[height=2.6em]{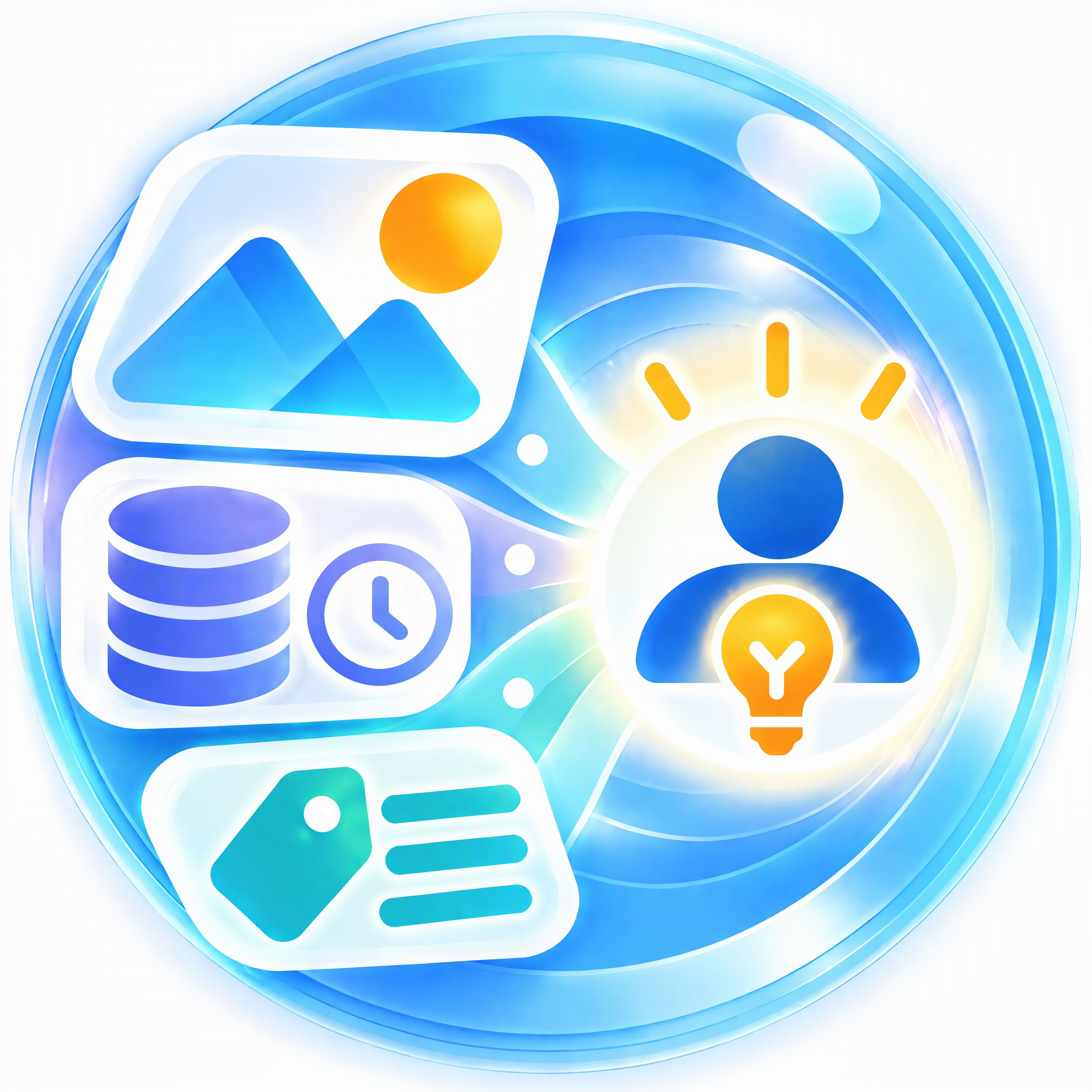}
  }{}
}

\title{\titleicon\hspace{0.15em}IntentLens: Grounding Underspecified Multimodal Queries for Recommendation via Tool-Augmented Reasoning}

\author{%
  Xiao Chen$^{1}$\thanks{Corresponding authors.},
  Dong Fang$^{2}$,
  Haitao Li$^{2}$\footnotemark[1],
  Qing Li$^{1}$\\
  $^{1}$The Hong Kong Polytechnic University,
  $^{2}$LightSpeed\\
  {\texttt{\{shawn.chen@connect.polyu.hk, loganli@lightspeed-studios.com\}}}
}

\begin{document}

\maketitle

\begin{abstract}

As intelligent personal assistants are increasingly deployed in real-world environments, recommendations move from passive preference matching to grounding user intent from underspecified multimodal evidence.
In these settings, users often provide only a reference image and a fuzzy natural-language query, leaving crucial visual attributes, personalized preferences, and item-side constraints implicit.
Existing recommenders either rely on historical interactions enriched with 
multimodal item features or presuppose fully specified textual requests, which are 
insufficient to resolve such queries for accurate item matching. To bridge this gap,
we present \textbf{IntentLens}, a tool-augmented recommendation framework that grounds 
underspecified multimodal queries into query-relevant evidence before user-item matching.
IntentLens adopts a \emph{ground-then-rank} design: a shared multimodal LLM 
orchestrates visual, user-memory, and item-attribute tools to (i) recover latent 
user intent from the image and language query, and (ii) enrich each candidate with 
fine-grained, query-conditioned evidence; a lightweight ranker then scores candidates 
over the grounded representations.
To support this new setting, we further construct two benchmarks with underspecified 
multimodal queries, \textsc{GoogleReview-MMTool} and \textsc{Yelp-MMTool}.
Comprehensive experiments on the constructed benchmarks show that IntentLens outperforms strong MLLM and retrieval baselines, and remains robust across unseen categories and datasets.

\end{abstract}

\section{Introduction}

Personalized recommendation is moving beyond passive preference matching toward \textbf{proactive intent understanding}~\cite{huang2025recommender,wang2024recmind,zhao2024let}. 
This trend is particularly relevant in physical settings, where camera-equipped wearables (e.g., Meta smart glasses and Apple Vision Pro) enable users to issue queries grounded in their immediate surroundings~\cite{yuan2026survey}. 
In these settings, users rarely fully specify their intent; instead, they often express their needs through a reference image paired with a fuzzy natural-language query. 
As a result, recommender agents must infer user intent from partial visual context and underspecified textual instructions, motivating tighter integration of multimodal reasoning into the recommendation process.

Despite the practical importance of this paradigm, existing recommendation systems~\cite{fan2022comprehensive,chen2023fairly,chen2025c2kd} are not well-suited to this setting. 
Existing methods mostly rely on simplifying assumptions: they either adopt a \textit{passive matching} approach, where multimodal features are fused mainly to enrich item representations guided by historical interactions~\cite{wu2022mm,liu2024rec,zhou2023mmrec,geng2023vip5}, or they assume \textit{fully specified textual queries} that explicitly state the user's constraints~\cite{kim2025lost,bao2025bi,jiang2025reclm,zhu2024visuallens}. 
In situated interactions, however, a multimodal query is often \textbf{underspecified}, providing only partial and ambiguous evidence of the user's latent intent. 
For example, a user may point their camera at a caf\'{e} dish and ask, \textit{``Something with this kind of vibe, but more affordable and better for a quick lunch nearby.''} 
Direct query--candidate matching can be unreliable in such cases, since important evidence, such as visual style, personal budget preferences, and item-specific attributes, may remain implicit or unavailable in the raw query.

Addressing underspecified multimodal queries requires recommender systems to go beyond passive matching and actively gather the evidence needed to infer user intent. In realistic recommendation scenarios, this evidence often comes from three sources: the query image, the user's interaction history, and item-side data. Accordingly, we equip the system with three tool categories: visual analysis tools that extract preference cues from the query image, memory retrieval tools that gather query-relevant information from user history, and item attribute tools that retrieve fine-grained metadata such as reviews to supplement sparse item descriptions.

In this work, we focus on recommendation systems that infer missing intent through tool-based evidence acquisition, thereby reducing the need for additional follow-up questions to users. We accordingly propose \textsc{IntentLens}, a two-stage \textit{ground-then-rank} framework. In Stage~1, a Multimodal Large Language Model (MLLM) selects and invokes query-side tools (i.e., visual and memory tools) as well as item-side tools to collect structured evidence, and then converts the resulting information into compact textual representations for the query and each candidate. In Stage~2, a lightweight dual-encoder ranker scores candidates against the grounded query representation, separating evidence acquisition from efficient ranking.

Pretrained MLLMs support basic tool invocation but are not trained to acquire evidence for personalized recommendations. Fine-tuning MLLMs is challenging in recommendation settings with many candidates, where rewards are sparse and effective planning requires exploring a large space of tool decisions. To better align tool decisions with downstream recommendation quality, we introduce a ranker-aligned Reinforcement Fine-Tuning (RFT) scheme that adapts the MLLM into a tool planner. We design a counterfactual ranking reward that compares each tool-using rollout with a no-tool baseline, making the value of acquired evidence explicit.  We optimize this objective with trajectory-level Group Relative Policy Optimization (GRPO), restricting gradient computation to policy-generated tokens by masking out environment-returned observations. Comprehensive experiments on our constructed benchmarks show promising recommendation quality of \textsc{IntentLens}.

Our main contributions are as follows:
\begin{itemize}
    \item We propose \textsc{IntentLens}, a two-stage framework that formulates multimodal recommendation under query underspecification as a problem of tool-augmented evidence grounding, separating intent grounding from candidate scoring for better scalability.
    \item We construct two new benchmarks, \textsc{Google Review-MMTool} and \textsc{Yelp-MMTool}, for evaluating recommendation with underspecified multimodal queries.
    \item We introduce a ranker-aligned RFT method for enhancing tool-using MLLMs, combining a counterfactual ranking reward with observation-masked GRPO to learn when and which tools improve recommendation quality.
\end{itemize}

\begin{figure}
    \centering
    \includegraphics[width=1\linewidth]{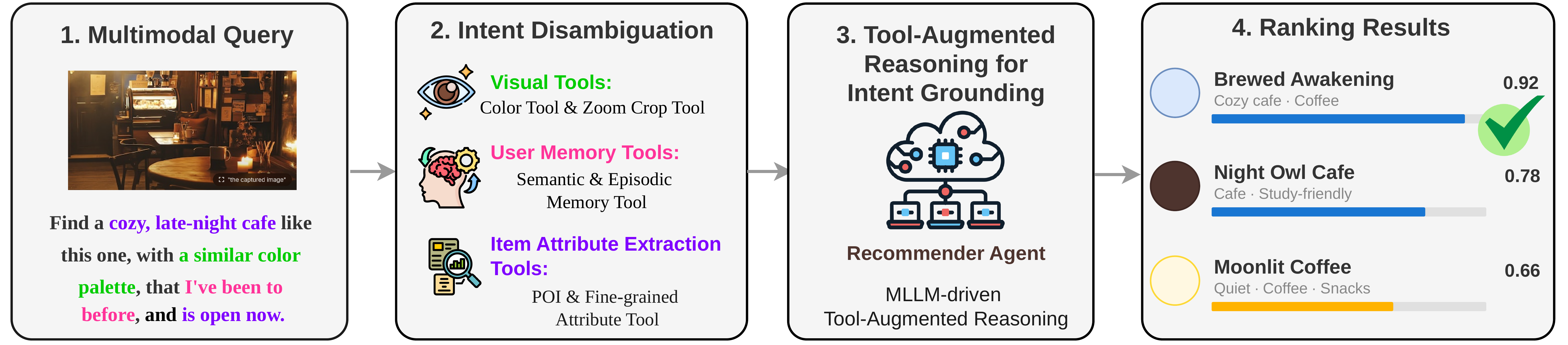}
    \caption{IntentLens leverages three types of tools to disambiguate underspecified multimodal queries, producing a clarified user intent and fine-grained candidate item attributes through MLLM-based tool-augmented reasoning, thereby enabling more accurate recommendations.}
    \label{fig:intro}
\end{figure}

\section{Related Work}

\paragraph{Multimodal Recommendation.}
Multimodal recommendation leverages heterogeneous signals such as text
and images to enrich user preference and item
representations~\cite{zhou2023mmrec,yan2023personalized,zhou2023comprehensive}.
Early approaches incorporate multimodal content as auxiliary item-side
features~\cite{he2016vbpr,wei2019mmgcn,zhang2021mining}, while subsequent
work develops graph-based architectures to improve representation
quality~\cite{guo2024lgmrec}.
These approaches primarily emphasize learning representations for
user--item matching~\cite{yuan2023go}.
Recent work introduces MLLMs into multimodal personalization and
recommendation~\cite{wei2024towards,ye2025harnessing,zhang2025notellm},
and VisualLens~\cite{zhu2024visuallens} further infers user preferences
from task-agnostic visual histories.
Our work complements representation learning with query-conditioned
evidence acquisition, using tools to ground underspecified multimodal
intent before ranking.

\paragraph{Underspecified Intent in Recommendation.}
Conversational recommender systems resolve ambiguous preferences through
multi-turn clarification~\cite{lei2020estimation,xu2021adapting,wang2022towards},
an approach also studied in information retrieval~\cite{zamani2020generating}.
History-based methods instead infer preferences from past
interactions~\cite{kang2018self,hou2022towards,zhou2020s3}.
Recent multimodal benchmarks examine related challenges:
RUMS~\cite{yan2025hidden} studies underspecified and misspecified
requests; MMA~\cite{wang2025mma} and MUCAR~\cite{wang2025mucar}
evaluate cross-modal ambiguity resolution; and
VAGUE~\cite{nam2025vague} examines how visual context supports
understanding implicit intent. 
We focus on single-turn personalized recommendation, using tools to
recover implicit preferences and item constraints without follow-up
dialogue.

\paragraph{Tool-Augmented Multimodal Reasoning.}
Visual ChatGPT~\cite{wu2023visual},
HuggingGPT~\cite{shen2023hugginggpt}, and
GPT4Tools~\cite{yang2023gpt4tools} enable language models to coordinate
external models and tools through prompting or instruction tuning.
MLLM-Tool~\cite{wang2025mllm} further incorporates multimodal inputs
into tool-agent learning.
Related benchmarks evaluate these capabilities across general-purpose
tasks: VisualToolBench~\cite{guo2025beyond} includes underspecified
requests requiring follow-up clarification, while
M$^3$-Bench~\cite{zhou2025m} evaluates multimodal tool orchestration
under the MCP protocol. RecToolBench~\cite{chen2026rectoolbench} further evaluates
recommendation-specific tool orchestration under fuzzy user instructions
in MCP-based environments.
Further discussion of related benchmarks is described in Appendix~\ref{sec:append_related}.

\begin{table}[t]
\centering
\scriptsize
\setlength{\tabcolsep}{4pt}
\renewcommand{\arraystretch}{1.06}
\caption{
Comparison of \textsc{IntentLens} with representative multimodal recommendation settings
and multimodal ambiguity/tool-use benchmarks.
\textbf{Image in Query}: the benchmark explicitly includes image input in the query.
\textbf{Underspecified Intent}: the primary source of under-specification in the benchmark;
$\circ$ indicates partial support.
\textbf{User Profile}: how user-specific information is accessed during inference.
\textbf{Tool-Augmented Intent Grounding}: whether correct inference requires external tools to ground evidence and recover the underlying intent.
}
\label{tab:bench_comparison}
\resizebox{\columnwidth}{!}{%
\begin{tabular}{l c >{\centering\arraybackslash}p{2.4cm} >{\centering\arraybackslash}p{2.4cm} c}
\toprule
\textbf{Benchmark}
& \makecell[c]{\textbf{Image in}\\\textbf{Query}}
& \makecell[c]{\textbf{Underspecified}\\\textbf{Intent}}
& \makecell[c]{\textbf{User Profile}}
& \makecell[c]{\textbf{Tool-Augmented}\\\textbf{Intent Grounding}} \\

\midrule
PC$^2$L~\cite{yan2023personalized}
  & \xmark & \xmark & Selected images & \xmark \\
COURIER~\cite{yang2025courier}
  & \xmark & \xmark & Multimodal embedding & \xmark \\
VisualLens~\cite{zhu2024visuallens}
  & \xmark & \xmark & Retrieved spectrum & \xmark \\
RectoolBench~\cite{chen2026rectoolbench}  & \xmark & Fuzzy task & \xmark & \cmark \\
Imagery as Inquiry~\cite{yoon2024imagery}
  & \cmark & Fuzzy preference & \xmark & \xmark \\
\midrule
ToolMMBench~\cite{wang2025mllm}
  & \cmark & Fuzzy task & \xmark & \cmark \\
VisualToolBench~\cite{guo2025beyond}
  & \cmark & Fuzzy task & \xmark & \cmark \\
M$^3$-Bench~\cite{zhou2025m}
  & \cmark & Fuzzy task & \xmark & \cmark \\
\midrule
\rowcolor{gray!12}
\textbf{\textit{IntentLens} (Ours)}
  & \cmark & Fuzzy preference +\ Fuzzy item constraints & Dual-level user memory & \cmark \\
\bottomrule
\end{tabular}%
}
\end{table}


\section{Multimodal Recommendation with Underspecified User Intent}                    
\subsection{Problem Definition}                                                   
  Given a user $u$ with historical interactions $\mathcal{H}_u$, a multimodal query $x=(I,q)$ consisting of a reference image $I$ and a natural-language request $q$, and a 
  candidate set $\mathcal{C}=\{c_i\}_{i=1}^{N}$ where each $c_i$ carries textual description $d_i$ and visual content $v_i$, the goal is to produce a ranking over
  $\mathcal{C}$ that aligns with the user's latent preferences.

  The key challenge is that the query provides only an incomplete and ambiguous specification of the user's true intent.
  We decompose the latent intent as $z_u = (z_u^{\mathrm{vis}},\, z_u^{\mathrm{hist}},\, z_u^{\mathrm{attr}})$, corresponding to:
  (i)~visual preferences inferred from the reference image, encoding perceptual cues that are difficult to verbalize;
  (ii)~personalized preferences reflected in historical interactions, capturing stable patterns or salient episodic signals;
  and (iii)~factual requirements on entity-level attributes that demand fine-grained matching on the candidate side.
  In this work, we address this problem from a \emph{tool-augmented reasoning} perspective: the system actively acquires evidence to disambiguate intent components and grounds its ranking on the enriched understanding.

  \subsection{Multimodal Tools for User Intent Disambiguation}

  We define a structured tool set aligned with the three components of $z_u$:

  \textbf{Visual tools} extract query-conditioned perceptual cues from the reference image, $\mathcal{E}^{\mathrm{vis}} = T_{\mathrm{vis}}(I, q)$, resolving underspecified
  visual preferences (e.g., color temperature) not stated in the text.

  \textbf{User memory tools} retrieve preference evidence from $\mathcal{H}_u$: $\mathcal{E}^{\mathrm{hist}} = T_{\mathrm{hist}}(I, q, \mathcal{H}_u)$. Following           
  A-Mem~\cite{xu2025mem}, we encode interactions into incrementally updated memory notes at two levels: episodic memories for interaction-specific contexts, and semantic
  memories for persistent preference patterns.                                                                                                                              
                  
  \textbf{Item attribute tools} provide fine-grained attribute evidence for candidate matching: $\mathcal{E}^{\mathrm{attr}} = T_{\mathrm{attr}}(\mathcal{C})$. These are
  deterministic extractors applied in a decoupled pass without MLLM reasoning. Detailed implementations of each tool are provided in Appendix~\ref{sec:tool_set}.

\section{The Proposed IntentLens Framework}
\begin{figure}
    \centering
    \includegraphics[width=0.9\linewidth]{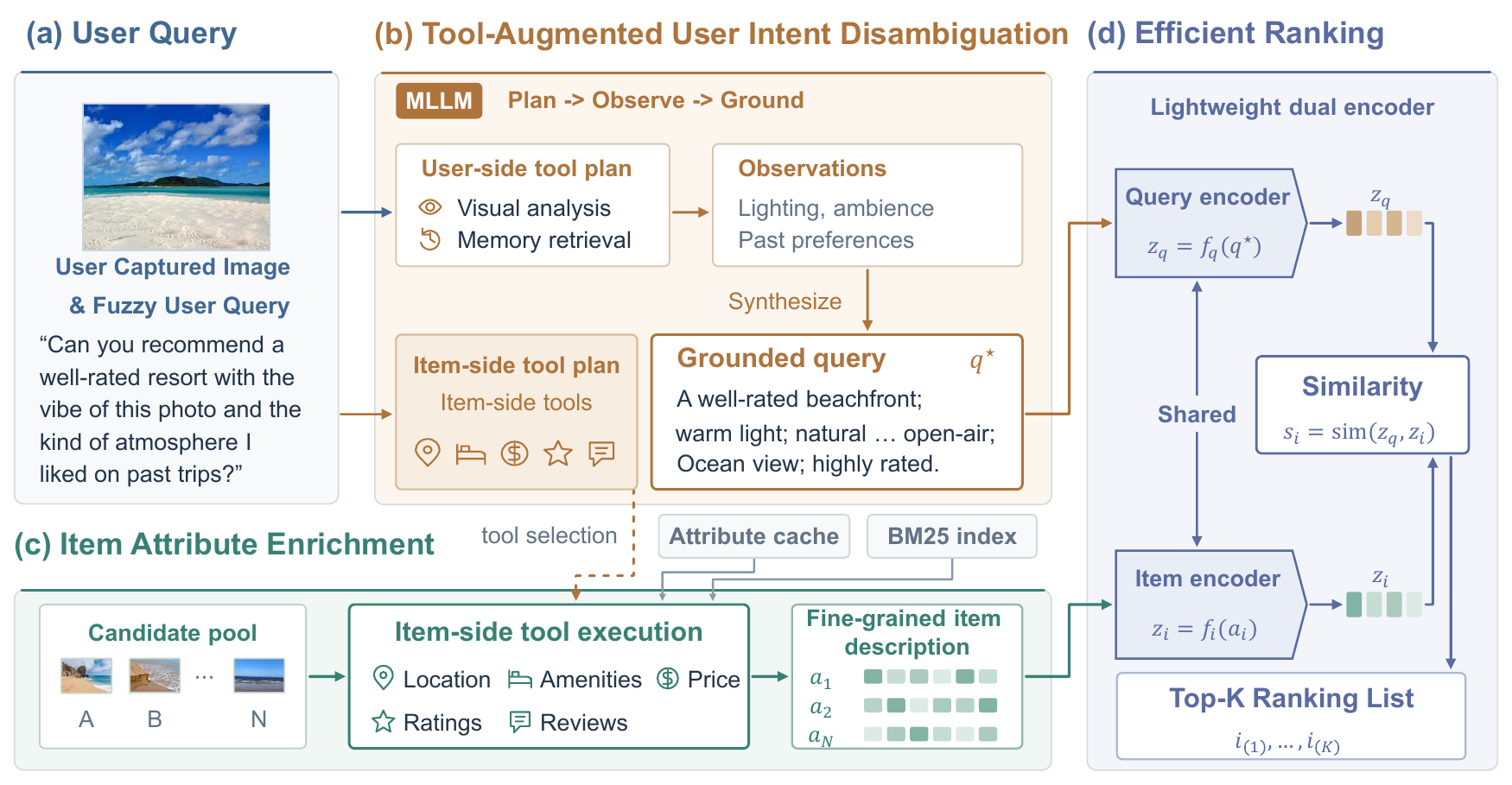}
    \caption{The inference pipeline of \textsc{IntentLens}. The MLLM first performs tool-augmented reasoning to produce a clarified query representation and select relevant item-side tools. The selected tools are then applied to the candidate pool in a decoupled pass, yielding fine-grained item representations. Finally, a lightweight dual-encoder ranker scores candidates against the grounded query to enable efficient recommendations. }
    \label{fig: Main_frm}
\end{figure}
We propose \textbf{IntentLens}, a two-stage framework for multimodal recommendation with underspecified user queries. The overall framework is shown in Figure~\ref{fig: Main_frm}.                                             
  In the first stage, an MLLM performs tool-augmented reasoning to disambiguate user intent and produces a grounded query representation. As part of this stage, the model predicts item-side tools relevant to the current query; these tools are then applied to the candidate pool in a decoupled pass, yielding fine-grained item      
  representations.     
  In the second stage, a lightweight dual-encoder ranker scores the grounded query against enriched candidate-item representations to enable efficient retrieval over large candidate pools.

  \subsection{Intent Grounding via Tool-Augmented Reasoning}
  \label{sec:grounding}

  Given a multimodal query $x=(I,q)$, IntentLens uses an MLLM to ground user intent from both visual and textual inputs.
  To keep online computation tractable, we separate user-side intent disambiguation from item-side attribute extraction.

  \paragraph{User Intent Disambiguation.}
  At inference time, the model resolves ambiguities in visual preference and user preference signals through user-side tools.
  We denote the user-side tool set by
  $\mathcal{T}_{\mathrm{user}} = \{T_{\mathrm{vis}}, T_{\mathrm{hist}}\}$,
  where $T_{\mathrm{vis}}$ extracts evidence from the query image and $T_{\mathrm{hist}}$ retrieves relevant signals from the user history $\mathcal{H}_u$.
  Rather than multi-turn sequential tool reasoning, we adopt a two-round procedure for efficiency.
  
  \textbf{i) In the first round}, the policy $\pi_\theta$ predicts an ordered tool plan, conditioned on the multimodal query (detailed prompt in
  Appendix~\ref{sec:appen_train_details}):
  \begin{equation}
  \mathcal{A}_{\mathrm{user}} \sim \pi_\theta\!\left(
\,\cdot \mid I, q , u;\, \mathcal{T}_{\mathrm{user}}
\right).
  \label{eq:tool-planning}
  \end{equation}
  Each action $a_k \in \mathcal{A}_{\mathrm{user}}$ is then executed by the external tool environment, producing observations
  \begin{equation}
  o_k = \mathtt{Execute}(a_k, I, q, \mathcal{H}_u), \quad \forall a_k \in \mathcal{A}_{\mathrm{user}},
  \qquad
  \mathcal{O}_q = \{o_k\}_{k=1}^{|\mathcal{A}_{\mathrm{user}}|}.
  \label{eq:tool-execution}
  \end{equation}

  \textbf{ii) In the second round}, observations $\mathcal{O}_q$ are appended to the context, and the policy synthesizes them into a grounded query representation:
  \begin{equation}
  \tilde{x}_q \leftarrow \pi_\theta\!\left(\cdot \mid I, q, \mathcal{O}_q\right).
  \label{eq:grounded-query}
  \end{equation}
  The candidate set $\mathcal{C}$ is excluded at this stage to avoid context expansion from large candidate pools.

  \paragraph{Decoupled Item Attribute Enrichment.}
$\pi_\theta$ also predicts an item-side tool set $\mathcal{T}_{\mathrm{item}} \subseteq \mathcal{T}_{\mathrm{all}}$ relevant to the current query. Unlike user-side tools, which require tool-augmented reasoning during inference, item-side tools are lightweight deterministic extractors.
At inference time, the selected tools expose cached item attributes
and review evidence without additional MLLM calls.
For each candidate, we combine its base description and attributes, and the
cached outputs associated with the selected item-side tools.
We also include precomputed candidate-image color features
corresponding to visual tools executed on the query side.
The resulting representation is
\begin{equation}
    \tilde{x}_i
    =
    \operatorname{Render}_{\mathrm{item}}
    \!\left(c_i,\mathcal{T}_{\mathrm{item}},\mathcal{O}_q\right),
    \qquad \forall\, c_i \in \mathcal{C},
    \label{eq:item-enrich}
\end{equation}
where $\mathcal{O}_q$ denotes the query-side tool observations,
and $\operatorname{Render}_{\mathrm{item}}$ assembles the available
fields in a fixed textual order.
This decoupling keeps online MLLM inference focused on query
understanding and tool selection, while item enrichment scales
linearly with the candidate pool size for a fixed tool inventory
and bounded field lengths.

\subsection{Reinforcement Fine-Tuning for Intent Grounding}
  \label{sec:rft}
             
While pretrained MLLMs support basic tool use, this alone does not guarantee that acquired evidence benefits recommendation quality.
In our setting, each rollout $\tau$ produced by $\pi_\theta$ contains a clarified query $q^{*}$ grounded in user-side evidence and an item-side tool set $\mathcal{T}_{\mathrm{item}}$ for candidate enrichment. To strengthen MLLM's tool reasoning ability, we apply reinforcement fine-tuning (RFT) with rewards tied to downstream ranking.

\paragraph{Counterfactual Ranking Reward.}
A frozen dual-encoder ranker $\hat{g}_{\phi}$
(Sec.~\ref{sec:rank}) evaluates each rollout over the full
candidate pool $\mathcal{C}$.
Given the relevant items $\mathcal{G}\subseteq\mathcal{C}$,
we define ranking utility as
$U(\tau)=1/\min_{c\in\mathcal{G}}\operatorname{rank}_{\tau}(c)$.
To measure the benefit of query-side evidence, we compare
$U(\tau)$ with two baselines: the same clarified query without
tool evidence, yielding $U_{\varnothing}(\tau)$, and the
original request without tool evidence, yielding
$U_{\mathrm{req}}(\tau)$.
All evaluations use identical candidate representations.
We define the evidence credit as
\begin{equation}
    \Delta_{\mathrm{cf}}(\tau)
    =
    \min\!\left\{
        U(\tau)-U_{\varnothing}(\tau),\,
        \left[U(\tau)-U_{\mathrm{req}}(\tau)\right]_{+}
    \right\},
    \label{eq:counterfactual-credit}
\end{equation}
where $[z]_{+}=\max(0,z)$.
Thus, positive credit requires improvement over both baselines.
The ranking reward is
\begin{equation}
    r_{\mathrm{rank}}(\tau)
    =
    U(\tau)
    +
    \beta\,\operatorname{clip}
    \!\left(\Delta_{\mathrm{cf}}(\tau),-\delta,\delta\right),
    \label{eq:ranking-reward}
\end{equation}
with $\beta=0.2$ and $\delta=0.1$.
The evidence bonus is zero when no usable evidence is returned.

\paragraph{Query-Side and Item-Side Tool Rewards.}
We reward accurate tool selection on both sides while penalizing
unnecessary tools. For each side $s\in\{q,i\}$, we define
\begin{equation}
    r_s(\tau)
    =
    \mathrm{F1}\!\left(
        \mathcal{T}_s(\tau),\mathcal{T}_s^{*}
    \right)
    -
    \gamma
    \frac{
        |\mathcal{T}_s(\tau)\setminus\mathcal{T}_s^{*}|
    }{
        |\mathcal{T}_s^{\mathrm{all}}|
    },
    \label{eq:tool-reward}
\end{equation}
where $\mathcal{T}_s^{*}$ is the target tool set,
$\mathcal{T}_s^{\mathrm{all}}$ is the full tool inventory for side $s$,
and $\gamma=0.3$.
Query-side selection counts only successfully executed tools
retained in the observations; optional tools outside the target
set are excluded from scoring.
We define $\mathrm{F1}(\varnothing,\varnothing)=1$.

In total, we combine ranking quality, tool selection, and format validity:
\begin{equation}
    r_{\mathrm{main}}(\tau)
    =
    r_{\mathrm{rank}}(\tau)
    +
    \lambda_q r_q(\tau)
    +
    \lambda_i r_i(\tau)
    +
    w_{\mathrm{fmt}}r_{\mathrm{fmt}}(\tau),
    \label{eq:total-reward}
\end{equation}
where $\lambda_q=\lambda_i=0.05$ weight the query-side and item-side
tool-selection rewards, and $w_{\mathrm{fmt}}=0.05$
weights the format reward.
The format reward $r_{\mathrm{fmt}}(\tau)$ equals one when
the first-round tool plan and second-round response satisfy
the required formats, and zero otherwise.
Ranking and tool-selection credit are masked out for
format-invalid rollouts; execution penalties are handled separately.

\paragraph{Trajectory-Level Masked GRPO.}
We use GRPO~\cite{shao2024deepseekmath} with losses restricted to policy-generated tokens.
Each rollout $\tau_i=(y_i^{(1)},\mathcal{O}_i,y_i^{(2)})$ contains two policy responses and intervening tool observations.
Let $\mathcal{P}_i$ denote the token positions in the two responses. Observations remain in the conditioning context but are excluded from both the policy and KL losses.
For a micro-batch $\mathcal{B}$ of rollouts, per instance with group-normalized advantages $\hat{A}_i = ({r(\tau_i)-\mathrm{mean}(\mathbf{r})})/{\mathrm{std}(\mathbf{r})}$, the objective is
\begin{equation}
\mathcal{J}_{\mathrm{MGRPO}}(\theta;\mathcal{B})
=
\frac{
    \sum_{i\in\mathcal{B}}\sum_{l\in\mathcal{P}_i}
    \ell\!\left(\rho_{i,l}(\theta),\hat{A}_i\right)
}{
    \sum_{i\in\mathcal{B}}|\mathcal{P}_i|
}
-
\mu\,
\widehat{\mathbb{D}}_{\mathrm{KL}}^{\mathcal{P}_{\mathcal{B}}}
\!\left[\pi_\theta\,\|\,\pi_{\mathrm{ref}}\right],
\label{eq:masked-grpo}
\end{equation}

where
$\rho_{i,l}(\theta)=
\pi_\theta(y_{i,l}\mid x_i,\tau_i^{<l})/
\pi_{\mathrm{old}}(y_{i,l}\mid x_i,\tau_i^{<l})$,
$\ell$ is the clipped policy surrogate, and $\mathcal{P}_i$
contains the policy-generated token positions from both rounds.
The objective averages over policy-generated tokens
across the batch.
We set rollout group size as $G{=}4$ for training efficiency.

  \subsection{Lightweight Ranking over Grounded Evidence}
  \label{sec:rank}

  After Stage~1, the query is grounded into $\tilde{x}_q$ and each candidate is enriched into $\tilde{x}_i$.
  Stage~2 delegates scoring to a dual-encoder ranker $g_\phi$ operating on these representations, with no further tool invocation.
  We pre-train $g_\phi$ on grounded representations produced by executing the ground-truth tool chains, then freeze it as $\hat{g}_\phi$ to serve as the stationary reward
  signal during RFT.

  \paragraph{Dual-Encoder Scorer.}
  We use a shared encoder to map the grounded query and enriched candidate descriptions into embeddings:
  \begin{equation}
\mathbf u_q=
\operatorname{Norm}\!\left(\operatorname{CLS}
(\operatorname{Enc}_\phi(\tilde x_q))\right),
\qquad
\mathbf v_i=
\operatorname{Norm}\!\left(\operatorname{CLS}
(\operatorname{Enc}_\phi(\tilde x_i))\right),
\qquad
s_i=\mathbf u_q^\top\mathbf v_i.
  \label{eq:rank}
  \end{equation}
where $\mathrm{Enc}_\phi$ is a shared BGE-M3 encoder backbone with CLS pooling and L2 normalization. Candidates are ranked by the inner product of the query and candidate embeddings.

\paragraph{Training Objective.}
  We train $g_\phi$ with the standard InfoNCE loss~\cite{oord2018representation}. For each query, positives are drawn from the ground-truth set
  $\mathcal{G}\subset\mathcal{C}$, and hard negatives are mined from same-category candidates ($\mathcal{N}\cap\mathcal{G}=\emptyset$) to improve representation discriminability.

  \section{Benchmark Construction}

\begin{table}[t]    
  \centering
  \caption{Hit rates and MRRs of \textsc{IntentLens} \textit{vs.} multiple baselines on Google Review-MMTool and Yelp-MMTool  $^* (\textit{p-value} \ll 0.05)$. Results for \textsc{IntentLens} are reported over $5$ runs. The best performance is highlighted in bold.}
  \label{tab:main_results}                                                       
  \resizebox{\columnwidth}{!}{%
  \begin{tabular}{lcccccccccc}
    \toprule           
   & & & \multicolumn{4}{c}{Google Review-MMTool} & \multicolumn{4}{c}{Yelp-MMTool} \\
  \cmidrule(lr){4-7} \cmidrule(lr){8-11}                                                
   & Modality & Size & Hit@1 & Hit@3 & Hit@10 & MRR & Hit@1 & Hit@3 & Hit@10 & MRR \\
  \midrule
   \multicolumn{11}{l}{\textit{(i) Basic baselines}} \\
  Random          & -   & -    & 0.0 & 10.3 & 27.6 & 9.9  & 2.8 & 7.0  & 18.3 & 9.2 \\
  Rank by Rating  & -   & -    & 0.0 & 3.5  & 31.0 & 9.2  & 2.8 & 8.5  & 31.0 & 11.0 \\
  Popularity      & -   & -    & 0.0 & 6.9  & 41.4 & 11.4 & 4.2 & 16.9 & 35.2 & 14.5 \\
  \midrule

  \multicolumn{11}{l}{\textit{(ii) Multimodal matching baselines}} \\
  CLIP Retrieval~\cite{radford2021learning}
                  & T+I & 428M & 3.5  & 13.8 & 24.1 & 12.8 & 14.1 & 26.8 & 47.9 & 24.6 \\
  BGE Retrieval~\cite{chen2024bge}
                  & T   & 335M & 13.8 & 27.6 & 37.9 & 25.7 & 19.7 & 36.6 & 57.8 & 32.2 \\
  \midrule

  \multicolumn{11}{l}{\textit{(iii) Zero-shot / Fine-tuned recommenders}} \\
  UniMP~\cite{wei2024towards}
                  & T+I & 3B   & 24.1 & 31.0 & 61.4 & 33.9 & 28.2 & \textbf{52.1} & 67.6 & 43.7 \\
  VIP5~\cite{geng2023vip5}
      & T+I & - & 3.5 & 24.1 & 51.7 & 19.6 & 4.2 & 14.1 & 33.8 & 14.4 \\
  Qwen2.5-VL (zero-shot)
      & T+I & 3B & 13.8 & 27.6 & 55.2 & 27.3 & 12.7 & 32.4 & 54.9 & 26.9 \\
  Qwen2.5-VL (zero-shot)
      & T+I & 7B & 13.8 & 31.0 & 58.6 & 28.0 & 12.7 & 22.5 & 56.3 & 25.4 \\
  \midrule

  \multicolumn{11}{l}{\textit{(iv) LLM/MLLM ranking}} \\
  GPT-5.4 
      & T   & - & 27.6 & 41.4 & 58.6 & 38.1 & 22.5 & 39.4 & 50.7 & 33.6 \\
  GPT-5.4 (w/ evidence)
      & T   & - & \textbf{31.0} & \textbf{44.8} & 55.2 & \textbf{41.6} & 31.0 & 43.7 & 67.6 & 41.5 \\
  GPT-4o
      & T+I & - & 30.9	& 33.6	& 40.4	& 36.8 & 31.0 & 45.1 & 66.2 & 41.4 \\
  \midrule

  \multicolumn{11}{l}{\textit{Our method}} \\
  \textsc{IntentLens} (3B)
                  & T+I & 3B   & 20.7 $\pm$ 2.0 & 41.4 $\pm$ 1.6 & 57.0 $\pm$ 1.8 & 34.5 $\pm$ 1.1 & 23.9 $\pm$ 0.9 & 32.4 $\pm$ 0.8 & 53.5 $\pm$ 0.6 & 33.5 $\pm$ 0.4 \\
  \textsc{IntentLens} (7B)
                  & T+I & 7B   & \textbf{31.0 $\pm$ 1.7} & \textbf{44.8 $\pm$ 1.4} & \textbf{62.1 $\pm$ 1.1} & 40.2 $\pm$ 1.2	 & \textbf{33.8 $\pm$ 1.4} & \textbf{52.1 $\pm$ 1.0} & \textbf{70.4 $\pm$ 0.8} & \textbf{46.7 $\pm$ 1.1} \\
  \midrule

  Oracle Ranker{*}
                  & T+I & -    & 37.9 & 41.4 & 65.5 & 44.4 & 43.7 & 53.5 & 76.1 & 53.7 \\
  \bottomrule
  \end{tabular}
  }
  \end{table}

  Existing recommendation benchmarks do not provide multimodal queries with underspecified user intent.                                                     Thus, we construct two benchmarks, \textsc{Google Review-MMTool} and \textsc{Yelp-MMTool}, from Google Local Review~\cite{li2022uctopic} and Yelp 2019~\cite{asghar2016yelp}. The benchmark statistics are detailed in Table~\ref{tab:dataset_stats}.

  \begin{table}[t]
  \centering
  \caption{Benchmark statistics for \textsc{Google Review-MMTool} and \textsc{Yelp-MMTool}.}
  \label{tab:dataset_stats}
  \setlength{\tabcolsep}{8pt}
  \begin{tabular}{lcccccc}
  \toprule
  Dataset & Users & Interact. & Queries & Categ. & Avg.\#GT & Avg.\#Cand. \\
  \midrule
  \textsc{GR-MMTool}   & 3{,}000 & 104{,}025 & 773 & 1{,}556 & 3.2 & 40 \\
  \textsc{Yelp-MMTool} & 2{,}315 & 94{,}184 & 1{,}727 & 497 & 2.0 & 63 \\
  \bottomrule
  \end{tabular}
  \end{table}

  \paragraph{User Behavior Logs and Memory.}
  For each user, reviews are sorted chronologically and linked to business metadata (name, category, description) and photos (user-uploaded in Google; store-profile images
  in Yelp).
  From these logs, we construct a dual-memory representation:
  (1)~\emph{episodic memory} storing individual interactions with associated text and images, and
  (2)~\emph{semantic memory} summarizing long-term preferences (favored categories, budget, atmosphere, distance tolerance) via LLM prompting over the user's prior review history (excluding target interactions).

  \paragraph{Query Synthesis.}
  Each query consists of a reference image (visual anchor) and an underspecified natural-language request that leaves key constraints implicit. Queries are organized into four complexity tiers by the tool categories required for resolution: (1)~visual only, (2)~visual + user memory, (3)~visual + item attribute, and (4)~visual + user memory + 
  item attribute. We use a three-stage forward-synthesis pipeline:
  (i)~grounded seeds are built by executing the corresponding tool combination on references retrieved via caption-similarity filtering from the user's history;            
  (ii)~an LLM verbalizes the tool evidence into an explicit intent description, then rewrites it into a natural underspecified query without exposing tool parameters;
  (iii)~automatic leak validation removes queries whose surface text reveals the target item or tool chain, and post-hoc human verification confirms that the intended reasoning path remains recoverable.

  \paragraph{Candidate Pool and Ground Truth.}
  For each query, we construct a candidate pool from nearby same-category businesses, containing positives, hard negatives from three sources (\emph{visual-similar},
  \emph{memory-conflicting}, \emph{context-conflicting}), and random distractors.                                                                                           
  Each candidate $c_i = (x_i, \mathcal{I}_i, a_i)$ carries text, an optional image set, and structured attributes.
  Ground truth is set-valued: a candidate is positive only if it jointly satisfies textual relevance, visual alignment, preference consistency, and item-side attribute constraints, determined by rule-based attribute filtering followed by LLM semantic verification.                                          Details of data synthesis are in Appendix~\ref{sec:append_bench}.

\paragraph{Human validation of benchmark quality.}
Three external annotators independently audited 400 queries, sampled
proportionally across T1--T4 within each benchmark (200 per benchmark).
Of these, 374 (93.5\%) met the majority-vote criteria for coherence,
naturalness, comprehensibility, and need for external evidence; dimension-wise
ordinal Krippendorff's $\alpha$ ranged from 0.67 to 0.74. Under label-blind
assessment, annotators confirmed 1,224 of 1,290 labeled positives (94.9\%;
$\alpha=0.68$). 
\section{Experiment}
\label{Experiments}

Based on the constructed benchmark, we conduct experiments to answer the following questions:
\textbf{Q1:} To what extent can tool-augmented intent grounding improve recommendations with underspecified multimodal queries?
\textbf{Q2:} How does each component of \textsc{IntentLens} contribute to recommendation quality?
\textbf{Q3:} Does \textsc{IntentLens} generalize across unseen categories, datasets and tool categories?

\subsection{Implementation and Baselines}
\textbf{Implementation Details.}
We implement \textsc{IntentLens} with two MLLM backbones, Qwen2.5-VL-3B and Qwen2.5-VL-7B~\cite{bai2025qwen3}. For RFT, we adopt the Verl framework with FSDP sharding over 4 H20 GPUs and optimize the policy using MGRPO. Rollouts are generated with vLLM. The reward model is a frozen dual-encoder ranker $\hat{g}_\phi$, trained on the training split and frozen during RFT.
The available tool set includes 5 visual tools, 3 user-memory tools, and 6 item-attribute tools. 

  
  
  
  
  
\textbf{Baselines.}  We evaluate five categories of baselines. For each evaluated query, all methods rank the same complete candidate pool.

(i) \textbf{Basic baselines:}
\textit{Random} uniformly shuffles the candidates;
\textit{Rating} ranks candidates by descending average star rating;
and \textit{Popularity} ranks candidates by descending review count.

(ii) \textbf{Multimodal retrieval baselines:}
\textit{CLIP Retrieval}~\cite{radford2021learning} uses CLIP ViT-L/14 to encode the reference image, textual request, and candidate descriptions. Candidates are ranked by the weighted average of image--candidate and text--candidate cosine similarities.
\textit{BGE Retrieval}~\cite{chen2024bge} concatenates the request with a textual caption of the reference image, encodes this query and candidate descriptions using BGE-large-en-v1.5, and ranks candidates by cosine similarity.

(iii)  \textbf{Zero-shot / Fine-tuned multimodal recommenders:}
We evaluate task-adapted implementations of \textit{UniMP}~\cite{wei2024towards} and \textit{VIP5}~\cite{geng2023vip5}, alongside \textit{Qwen2.5-VL-3B} and \textit{Qwen2.5-VL-7B} in a zero-shot setting.
Our UniMP-style implementation uses Qwen2.5-VL-3B with the reference image, user request, and cleaned review history.
Our VIP5-style implementation augments T5-base~\cite{raffel2020exploring} with visual prefix tokens obtained by projecting frozen CLIP image features.
The Qwen~\cite{bai2025qwen3} baselines rank candidates without user history or tool use.

(iv) \textbf{LLM/MLLM ranking:}
We evaluate three ranking baselines: text-only \textit{GPT-5.4}, multimodal \textit{GPT-4o}~\cite{hurst2024gpt}, and tool-augmented \textit{GPT-5.4}.
The first two directly rank candidates using the request and candidate metadata, with GPT-4o additionally receiving the reference image.
The tool-augmented variant ranks candidates using the clarified query, query-side tool outputs, and candidate-side evidence generated by the fixed IntentLens policy.

(v) \textbf{Oracle Ranker:}
We evaluate frozen BGE-M3~\cite{chen2024bge} with the ground-truth clarified query, annotated query-side tool evidence, and annotated item-side tool selections.
This isolates the effect of annotated versus policy-generated inputs under the same ranker.

\noindent\textbf{Evaluation Metrics.}
  We report two ranking metrics.
  \textbf{Hit@$k$} measures whether at least one relevant item appears in the top-$k$ positions:
  $\mathrm{Hit@}k
  =
  \frac{1}{N}\sum_{i=1}^{N}
  \mathbb{I}\!\bigl[\mathrm{rank}(r_i)\le k\bigr]$,
  $\mathrm{rank}(r_i)$ is the position of the highest-ranked relevant item for query $i$.
  We report Hit@1, Hit@3 and Hit@10.
  \textbf{MRR} averages the reciprocal of this rank as
  $\mathrm{MRR}
  =
  \frac{1}{N}\sum_{i=1}^{N}
  \frac{1}{\mathrm{rank}(r_i)}$.

  \begin{table}[t]                                         \centering      
  \small                 
  \setlength{\tabcolsep}{2pt}
  \renewcommand{\arraystretch}{1.0}  
  \caption{Ablation study for analyzing different components of \textsc{IntentLens} Model on Google Review-MMTool. (a)~Tool group ablations. (b)~Training objective ablations.}       
  \label{tab:ablation}                                                           
  \begin{minipage}[t]{0.48\columnwidth}                                          
  \centering
  \subcaption{Tool categories}\label{tab:ablation_tool}
 \begin{tabular}{l cccc}
  \toprule
   Variant & Hit@1 & Hit@3 & Hit@10 & MRR \\                                     
  \midrule
   \textsc{IntentLens} (7B) & \textbf{31.0} & \textbf{44.8} & \textbf{62.1} & \textbf{40.2} \\                                                                           
  \midrule        
   w/o visual & 24.1 & 37.9 & 58.6 & 35.7 \\
   w/o memory & 26.3 & 37.9 & 59.6 & 38.1 \\
   w/o item & 27.6 & 41.4 & 58.4 & 38.0 \\
   w/o grounding & 21.9 & 36.8 & 51.7 & 33.5 \\
  \bottomrule
  \end{tabular}
  \end{minipage}%
  \hfill
  \begin{minipage}[t]{0.48\columnwidth}
  \centering
  \subcaption{Training objectives}\label{tab:ablation_train}
  \begin{tabular}{l cccc}
  \toprule
   Variant & Hit@1 & Hit@3 & Hit@10 & MRR \\
  \midrule
      \textsc{IntentLens} (7B) & \textbf{31.0} & \textbf{44.8} & \textbf{62.1} & \textbf{40.2} \\
  \midrule
   w/o RFT & 24.1 & 33.3 & 58.6 & 33.1 \\
  w/o counterfactual & 24.1 & 36.1 & 58.6 & 35.0 \\
  w/o tool reward & 24.1 & 37.9 & 62.1 & 35.5 \\
  w/o obs.\ masking & 28.9 & 38.6 & 60.5 & 35.1 \\
  \bottomrule
  \end{tabular}
  \end{minipage}
  \end{table}    

\subsection{Main Results}

\textbf{Comparison with retrieval and zero-shot baselines.}
Table~\ref{tab:main_results} shows that \textsc{IntentLens} (7B) outperforms CLIP, BGE, and both zero-shot Qwen2.5-VL variants across the board. 
These gains demonstrate the necessity of incorporating external, query-relevant evidence rather than relying solely on direct query--candidate similarity for underspecified multimodal recommendation.

\textbf{Comparison with competitive baselines.}
Compared to other non-oracle approaches, \textsc{IntentLens} (7B) sets the state of the art on most metrics: it achieves the highest Hit@10 on Google Review-MMTool, and tops Hit@1, Hit@10, and MRR on Yelp-MMTool. 
Notably, these results rely on pairing tool augmentation with the 7B backbone, indicating that tool use becomes fully effective only when supported by sufficient model capacity.

\textbf{Model scaling and oracle gap.}
Upgrading the backbone from 3B to 7B consistently benefits \textsc{IntentLens}, lifting MRR from 34.5 to 40.2 on Google Review-MMTool and from 33.5 to 46.7 on Yelp-MMTool. 
While scaling yields pronounced gains, particularly on Yelp-MMTool, the 7B model still lags behind the Oracle Ranker by 4.2 and 7.0 MRR points on the respective datasets, indicating remaining room to close the grounding gap.

\subsection{Ablation Study}
We evaluate the contributions of individual components of
\textsc{IntentLens} on Google Review-MMTool.
Results are reported in Table~\ref{tab:ablation}.

\noindent\textbf{Tool and grounding ablations.}
We consider four variants:
(i)~\textbf{w/o visual tools} disables visual tools;
(ii)~\textbf{w/o memory tools} removes user-memory retrieval tools;
(iii)~\textbf{w/o item tools} replaces tool-augmented candidate
representations with raw descriptions;
(iv)~\textbf{w/o query grounding} bypasses the entire query-side
tool pipeline.
Removing query grounding causes the largest decline across all
reported metrics among these variants, supporting its role in
resolving underspecified requests before ranking.
Removing visual, memory, or item tools also reduces all reported
metrics, indicating that both query-side and candidate-side
evidence contribute to recommendation quality.
Among the individual tool categories, removing visual tools
produces the largest MRR decline.

\noindent\textbf{Training objective ablations.}
We consider four additional variants:
(v)~\textbf{w/o RFT} uses only the SFT-initialized planner;
(vi)~\textbf{w/o counterfactual baseline} removes the no-tool
reference from the reward;
(vii)~\textbf{w/o item-side tool reward} drops the auxiliary
tool-selection reward;
(viii)~\textbf{w/o observation masking} includes tool-returned
observation tokens in both the policy and KL losses, while
retaining them in the context as in the full model.
Removing RFT causes the largest MRR decline among the training
variants, showing that reinforcement fine-tuning improves
recommendation quality beyond SFT initialization.
Removing the counterfactual baseline reduces all reported metrics.
Removing the item-side tool reward lowers Hit@1, Hit@3, and MRR, suggesting that this reward
helps place relevant candidates earlier in the ranking.
Disabling observation masking also reduces MRR,
supporting the use of loss masking for tool-returned tokens
rather than treating them as policy-generated outputs.

\noindent\textbf{Query representation control.}
On Google Review-MMTool, we hold the dual encoder, candidate pool,
and candidate-side evidence fixed and vary only the query representation.
Raw, predicted-grounded, and oracle-grounded queries yield MRRs
of 33.6\%, 40.2\%, and 44.4\%, respectively.
This controlled comparison isolates the contribution of query
grounding from changes to the ranker or candidate-side evidence.

\noindent\textbf{Alternative-ranker transfer.}
Using frozen \texttt{all-MiniLM-L6-v2}, neither tuned on our benchmark
nor used during RFT, we compare cached query and candidate evidence
from the pre-RFT instruction-base model and the RFT planner.
With candidate pools fixed, RFT improves MRR by 4.6 percentage points, indicating transfer beyond the original reward ranker.

\noindent\textbf{Human-judged recommendation quality.}
Three external annotators blindly compare randomized top-three
recommendation lists on 100 stratified queries, 50 per benchmark,
without system identities, scores, or automatic labels.
By majority vote, \textsc{IntentLens} is preferred on 89\% of queries
versus 6\% for BGE Retrieval, providing label-independent
evidence of recommendation quality.

\noindent\textbf{Human-written query transfer.}
We further conduct a user study to evaluate model robustness on human-authored queries. 
Given only the images and task briefs (with synthetic queries, candidates, and labels hidden), participants drafted 100 requests, of which 78 passed quality verification by two validators. 
Keeping the candidate pools fixed and applying no retraining or prompt tuning, \textsc{IntentLens}-7B still outperforms BGE by 7.3 MRR points on these human queries, confirming that its gains generalize well beyond synthetic query patterns.


\subsection{Generalization and Efficiency Analysis}
We evaluate \textsc{IntentLens} under unseen-category, cross-dataset,
and unseen-tool settings, as shown in Table~\ref{tab:gen_eff}(a).
MRR decreases under all three settings, with cross-dataset transfer
producing the largest decline on both benchmarks.
Performance remains closer to the in-domain results under unseen
categories and tools, suggesting that \textsc{IntentLens} can
generalize beyond the categories and tool configurations encountered
during training.
The larger cross-dataset gap indicates that transfer across platforms
remains more challenging in our evaluation.

Table~\ref{tab:gen_eff}(b) reports end-to-end latency of \textsc{IntentLens}-7B. The full pipeline processes a query in approximately 4 seconds, with Stage~1 (MLLM rollout and tool execution) dominating at 3.5 s and Stage~2 (dual-encoder scoring over candidates) adding only 515 ms. This two-stage design keeps inference speed tolerable for recommendation scenarios while preserving the utility of tool-augmented reasoning. 
We record Item Tool F1 and user-side Tool F1 alongside the Tool Call Ratio. As training progresses, both F1 scores rise steadily, while the Tool Call frequency decreases. This pattern indicates that the policy learns not only to select more appropriate tools but also to suppress redundant invocations, achieving more selective and precise tool planning.

\begin{table}[t]
  \centering
  \caption{(a)~Generalization analysis under different transfer setups. \textbf{Unseen Cat.}: hold out item categories during training. \textbf{Cross-Dataset}: train on the
   other dataset. \textbf{Unseen Tools}: remove a subset of tools from training.
  (b)~Efficiency analysis on a single NVIDIA H20 GPU. Stage~1: MLLM rollout for tool planning; Stage~2: frozen dual-encoder scoring.}                                    
   \label{tab:gen_eff}
  \begin{minipage}[t][3cm][t]{0.44\columnwidth}
  \centering
  \subcaption{Generalization Analysis}\label{tab:generalization}
  \small
  \setlength{\tabcolsep}{3pt}                                                                                  
  \begin{tabular}{l cccc}
  \toprule                                                                                           
  \textbf{Test Set} & In-Dom. & Uns.\ Cat. & Cross & Uns.\ Tool \\
  \midrule
  Google & 40.2 & 36.2 & 34.7 & 35.8 \\
  Yelp   & 46.7 & 43.5 & 38.9 & 44.1 \\
  \bottomrule
  \end{tabular}
  \end{minipage}%
  \hfill
  \begin{minipage}[t][2.2cm][t]{0.53\columnwidth}
  \centering
  \subcaption{Efficiency Analysis}\label{tab:efficiency}
  \small
  \setlength{\tabcolsep}{3pt}
  \begin{tabular}{l ccc}
  \toprule
  \textbf{Stage} & Latency (ms) & \#MLLM Calls & \#Tool Calls \\
  \midrule
  1  & 3,590 & 2 & 3--5 \\
  2  & 515   & 0 & 0    \\
  \bottomrule
  \end{tabular}
  \end{minipage}
\vspace{-0.5cm}
  \end{table}

\section{Conclusion and Limitation}
In this paper, we presented \textsc{IntentLens}, a two-stage framework that formulates multimodal recommendation under query underspecification as a tool-augmented reasoning problem. The proposed ranker-aligned reinforcement fine-tuning scheme, combining counterfactual ranking reward with observation-masked GRPO, enables the MLLM to learn selective and effective tool use tied to downstream recommendation quality. Experiments on our constructed Google Review-MMTool and Yelp-MMTool demonstrate consistent improvements against most baselines. Our findings are constrained by constructed benchmarks and single-turn evaluation, with limited applicability to real-world multi-turn interactions and cross-domain transfer, alongside practical deployment bottlenecks and privacy challenges tied to user and visual data.

\section*{Acknowledgments}
The experimental part of this work was partially supported by The Centre for Large AI Models (CLAIM) of The Hong Kong Polytechnic University. This work was also supported by The Hong Kong Polytechnic University through the Research Centre for Data Science and Artificial Intelligence (Project No. 4-CE42).

\clearpage

\bibliographystyle{unsrt}
\bibliography{references} 

\clearpage
\appendix


\section{More Related Work}
\label{sec:append_related}
\subsection{Underspecified Query Handling in MLLMs}

Handling underspecified or ambiguous user queries is a well-recognized challenge in information retrieval and dialogue systems.
\textbf{RUMS}~\cite{yan2025hidden} specifically diagnosing underspecified and misspecified scenarios for MLLMs across four implicit reasoning challenges: object absence, referential ambiguity, factual contradiction, and goal infeasibility.
A key finding is that clarification is the most effective inference-time intervention.
\textbf{MMA Benchmark}~\cite{wang2025mma} evaluates whether MLLMs can use images to disambiguate textually ambiguous queries (lexical, syntactic, and semantic), finding that MLLMs achieve only 53.22\% accuracy on paired-image disambiguation (vs. 88.97\% for humans).
\textbf{VAGUE}~\cite{nam2025vague} tests whether VLMs can leverage visual context to infer the speaker's implicit intent from indirect expressions (\emph{e.g.}, ``turn it off'' in a scene with multiple computers), revealing a $\sim$20\% gap between VLMs and humans.
\textbf{MUCAR}~\cite{wang2025mucar} further extends this to multilingual cross-modal ambiguity resolution, where both image and text are simultaneously ambiguous and must disambiguate each other.

\subsection{Multimodal Tool Calling and Reasoning}
Early work on tool-augmented language models primarily focused on text-only interfaces. Systems such as \textbf{Visual ChatGPT}~\cite{wu2023visual}, \textbf{HuggingGPT}~\cite{shen2023hugginggpt}, and \textbf{GPT4Tools}~\cite{yang2023gpt4tools} enable LLMs to invoke external tools through prompt-based or instruction tuning paradigms. However, these approaches rely on \emph{purely textual queries}, making tool selection brittle when the user intent heavily depends on visual context.
Recent advances have extended tool use to multimodal settings. \textbf{MLLM-Tool}~\cite{wang2025mllm} and subsequent works incorporate visual perception and improve tool invocation through supervised and reinforcement learning. 
Despite these impressive advances, existing methods primarily focus on \emph{tool selection accuracy or execution correctness} in general-purpose tasks, rather than reasoning over \emph{underspecified, preference-driven user intent} in real-world recommendation scenarios. Our framework bridges this gap by repurposing tool-augmented MLLMs as an intent planner. Instead of merely executing isolated commands, our planner orchestrates visual, memory, and item tools to gather explicit evidence, translating ambiguous multimodal intent into a structured textual interface that a lightweight recommender can efficiently process. 

\subsection{Benchmarks for Multimodal Tool Calling}

Several benchmarks have been proposed to evaluate MLLM tool-calling capabilities.
\textbf{VisualToolBench}~\cite{guo2025beyond} introduces 1,204 open-ended visual tasks across five domains (STEM, Medical, Finance, Sports, Generalist), covering six tool types including image processing APIs, Python interpreters, and web search.
Notably, it includes a \emph{Follow-up Test} category for multi-turn settings where the model must proactively ask clarifying questions when the query is underspecified.
\textbf{M$^3$-Bench}~\cite{zhou2025m} is the first benchmark under the MCP protocol, covering 28 MCP servers, 231 tools, and 208 evaluation trajectories, with fuzzy task descriptions designed to stress-test tool argument fidelity and structural consistency.

\section{Benchmark Details}
\label{sec:append_bench}

\subsection{Tool Set}
\label{sec:tool_set}

\textbf{Color-based visual tools} extract quantitative attributes such as color distribution, brightness, and saturation, which serve as proxies for high-level semantics like ambiance and style. For example, a color histogram captures the distribution of pixel intensities across color channels, enabling precise estimation of brightness and saturation levels.

These attributes are particularly critical for resolving relative and underspecified user queries, where the required reasoning involves comparing distributions rather than recognizing objects.

For instance, a query such as “find a place with a similar vibe but brighter and less heavy” requires analyzing the brightness distribution (via a color histogram) to quantify what “brighter” means, while a request like “something with a similar color palette but more neutral tones” relies on dominant color extraction to compare color composition across images. Similarly, queries such as “not too dark” or “more vibrant atmosphere” depend on estimating brightness and saturation levels, which cannot be reliably inferred from holistic VLM embeddings alone.

In addition, \textbf{spatial manipulation tools} such as crop and zoom-in are essential when the relevant visual signal is localized.
For example, given a query like “I like the plating style of this dish—recommend something similar nearby,” the model must first isolate the dish region from irrelevant background (e.g., table, people, lighting) via cropping before extracting visual attributes. Likewise, a request such as “something like this, but focusing on the texture/details” requires zooming into fine-grained regions (e.g., food texture or garnish), which cannot be explicitly controlled within a single forward pass of a VLM.

Together, these examples highlight that such queries require explicit region selection and quantitative attribute comparison, rather than holistic semantic understanding—capabilities that are not natively supported by VLM representations.

\textbf{User Memory Tools}.  To support users with hundreds of reviews and deployments with tens of thousands of users, we adopt a normalized storage design: review-level memory notes are linked to de-duplicated business profiles via gmap\_id, while extracted preference signals and user-level summaries are maintained in separate indexed structures. This design preserves interpretability at the logical level while keeping storage overhead and retrieval latency manageable at scale.

We expose the user memory bank through a small set of retrieval interfaces with complementary roles. RetrieveUserMemory performs query-conditioned retrieval over review-grounded episodic notes, GetUserProfileSummary returns an aggregated user-level preference prior, and GetMemoryNoteDetails supports selective expansion of retrieved notes when more evidence is needed. This design separates global preference summarization from local episodic evidence, while keeping the retrieval API compact and token-efficient.

\begin{table}[t]
\centering
\footnotesize
\setlength{\tabcolsep}{4pt}
\caption{Tools and their roles in resolving underspecified multimodal recommendation queries.}
\label{tab:visual_tools}
\begin{tabular}{p{2cm} p{3.3cm} p{4.1cm} p{3.6cm}}
\toprule
\multicolumn{1}{c}{\textbf{Tool Name}} & 
\multicolumn{1}{c}{\textbf{Tool Description}} & 
\multicolumn{1}{c}{\textbf{Query Example}} & 
\multicolumn{1}{c}{\textbf{Design Rationale}} \\
\midrule
\multicolumn{4}{c}{\textit{Visual Evidence Tools}} \\
\midrule
{Crop}
& Extracts a region of interest from the query image.
& \textit{I’m referring to the dish in the corner of this image, find restaurants that serve something similar.}
& Isolate the target dish from a cluttered scene before visual comparison. \\[4pt]
{Zoom In}
& Scales up a specific region for detail inspection.
& \textit{Find places with a similar style to the decor details in this café.}
& Fine-grained texture details are more salient via zooming. \\[4pt]
{Dominant Color}
& Extracts the top-$k$ perceptually dominant colors via $k$-means clustering.
& \textit{Find venues with a similar warm, earthy color palette.}
& Quantifies vague descriptors into a comparable color signature. \\[4pt]
{Color Histogram}
& Computes pixel-level color distribution across channels.
& \textit{Find a restaurant whose interior has a similar warm and soft color distribution.}
& Full color distribution distinguishes genuinely warm-toned interiors from others. \\[4pt]
{Color Temperature}
& Estimates correlated color temperature in Kelvin.
& \textit{I prefer a place with cozy lighting like this suitable for an evening date.}
& ``Cozy lighting'' is subjective; CCT gives an objective, matchable measure of warmth. \\
\midrule
\multicolumn{4}{c}{\textit{User Memory Retrieval Tools}} \\
\midrule
GetUserProfile & Get user's long-term preference profile & \textit{Recommend a place that generally fits my taste.} & Provides a long-term prior when the current query is too vague to infer preferences. \\
RetrieveMemory & Retrieve relevant historical reviews by structured slots  & \textit{I really liked that brunch place I went to before, find something similar} & Grounds fuzzy queries via matched past interactions.\\
GetDetails & expand the full content of specific memory notes  & \textit{Remember that spot with the amazing ramen I reviewed? I want something similar} & Resolves ambiguity by exposing fine-grained evidence from a retrieved memory note.
\\
\midrule
  \multicolumn{4}{c}{\textit{Item Evidence Retrieval Tools }} \\
  \midrule
  GetItemProfile & Retrieve the detailed profile of a candidate item (category, detailed description, amenities) & \textit{What kind of place is this exactly?} & Provides full item
  identity beyond name and category. \\
  \addlinespace
  GetReviewInsights & Extract review-based insights and sentiment for an item & \textit{What do people say about the atmosphere here?} & Surfaces experiential attributes
  not captured in metadata. \\                                                   
  \addlinespace
  GetRating & Query rating and review count for quality comparison & \textit{Find a place rated higher than Restaurant X.} & Supports relative quality filtering across     
  candidates. \\  
  \addlinespace
  GetPrice & Query price tier for cost comparison & \textit{Find somewhere cheaper than Restaurant X.} & Supports budget-aware filtering across candidates. \\
  \addlinespace
  GetHours & Query opening hours for availability check & \textit{Find a place open later than Restaurant X.} & Filters candidates by temporal availability. \\
  \addlinespace
  SearchByCondition & Search candidates matching specific attribute criteria & \textit{Find places with outdoor seating nearby.} & Enables multi-attribute filtering over
  the candidate pool. \\
\bottomrule
\end{tabular}
\end{table}

\begin{algorithm}[t]
\caption{Review-Grounded Memory Bank Construction}
\label{alg:memory_construction}
\begin{algorithmic}[1]
\Require User review history $\mathcal{R}_u = \{(r_i, \mathcal{P}_i, b_i)\}_{i=1}^{N}$ where $r_i$ is review text, $\mathcal{P}_i$ is associated photos, $b_i$ is business metadata; L1 visual tool set $\mathcal{T}_{\text{vis}}$; Teacher MLLM $\mathcal{M}$
\Ensure Memory Bank $\mathcal{B}_u = \{m_1, m_2, \ldots, m_N\}$
\Statex
\For{each review $(r_i, \mathcal{P}_i, b_i) \in \mathcal{R}_u$}
    \Statex \Comment{\textit{Step 1: Extract visual features via L1 tools}}
    \State $\mathbf{v}_i \gets \emptyset$
    \For{each photo $p \in \mathcal{P}_i$}
        \State $\mathbf{v}_i[p] \gets \{t(p) \mid t \in \mathcal{T}_{\text{vis}}\}$ \Comment{Run all visual tools}
    \EndFor
    \Statex \Comment{\textit{Step 2: Extract preference signals from review text}}
    \State $\mathbf{s}_i \gets \mathcal{M}(r_i, b_i, p_{\text{extract}})$ \Comment{Extract sentiment, aspects, keywords, occasion}
    \Statex \Comment{\textit{Step 3: Generate semantic tags for indexing}}
    \State $\tau_i \gets \mathcal{M}(\mathbf{v}_i, \mathbf{s}_i, b_i.\text{MISC}, p_{\text{tag}})$
    \Statex \Comment{\textit{Step 4: Construct atomic memory note}}
    \State $m_i \gets \textsc{MemoryNote}(r_i, b_i, \mathbf{v}_i, \mathbf{s}_i, \tau_i)$
    \State $\mathcal{B}_u \gets \mathcal{B}_u \cup \{m_i\}$
\EndFor
\Statex \Comment{\textit{Step 5: Build retrieval indices}}
\State $\mathcal{I}_{\text{vis}} \gets \textsc{BuildVisualIndex}(\{m_i.\mathbf{v}\})$ \Comment{Color temp / dominant color index}
\State $\mathcal{I}_{\text{text}} \gets \textsc{BuildTextIndex}(\{m_i.\mathbf{s}\})$ \Comment{Preference keyword embeddings}
\State $\mathcal{I}_{\text{tag}} \gets \textsc{BuildTagIndex}(\{m_i.\tau\})$ \Comment{Tag → note\_id mapping}
\State \Return $\mathcal{B}_u, \mathcal{I}_{\text{vis}}, \mathcal{I}_{\text{text}}, \mathcal{I}_{\text{tag}}$
\end{algorithmic}
\end{algorithm}

\subsection{Data Synthesis Implementation Details}
We synthesize four levels of underspecified user queries, where difficulty is defined by the number of tools required to resolve the query.

\paragraph{T1: Visual Venue Recommendation.}                                                                                                                              
T1 targets visual venue recommendation, where a user submits a venue photograph and requests places with a similar visual style. Instances are constructed through a three-stage forward synthesis pipeline.

  \textbf{Stage 1 (Seed Construction).}
  For each venue, we sample a reference image and retrieve candidate venues from the same city and category. Candidates are filtered by caption similarity (computed via    
  synonym-expanded token overlap with scene-type bonuses), a formality guard that removes cross-tier pairs (e.g., fast-casual vs. upscale), and a visual-domain compatibility check that ensures both images belong to the same scene type (food, outdoor, interior, etc.). Document-like images (menus, signs) are excluded.

  \textbf{Stage 2 (Query Generation).}
  A vision-language model (Qwen2.5-VL-7B) analyzes each reference image and produces a natural-language request describing abstract visual qualities (atmosphere,
  lighting, color tone, style) without revealing the venue identity. A text LLM   (GPT-5.1) then verifies coherence and finalizes the query.

  \textbf{Stage 3 (Ground-Truth Validation).}
  Visual tool verification determines the final positive set: we reject pairs with inconsistent color-temperature tiers (absolute difference $>$1500K or opposite warmth
  polarity) and filter out document-type images. All candidate venues passing these checks form the ground-truth set. Each T1 instance is a triple of (reference image,     
  natural-language query, ground-truth venue set).

\paragraph{T2: User Memory + Visual Tools.}

  T2 introduces queries that require joint reasoning over personal memory and visual signals. Each instance involves three items: a visual reference providing the image anchor, an optional episodic reference anchoring a past experience, and a target item for recommendation.
  
  \textbf{Seed Construction.}
  We filter users with sufficient history (${\geq}20$ interactions, ${\geq}3$ items with scene photos). For each user, the target is the most recent item with valid scene  
  captions. The visual reference is selected via caption similarity filtering and visual-domain compatibility, followed by color-temperature
  consistency validation to ensure visual tools produce meaningful signals.

  \textbf{Tool-Chain Construction.}
  Two memory retrieval modes alternate deterministically: \emph{profile mode} uses a single \texttt{GetUserProfileSummary} call, while \emph{episode mode} follows a
  three-step chain (\texttt{GetUserProfileSummary} $\to$ \texttt{RetrieveUserMemory} $\to$ \texttt{GetMemoryNoteDetails}).
  The visual tool chain is selected by an LLM (GPT-5.1) conditioned on the scene caption: simple scenes use a single-layer tool (e.g., \texttt{dominant\_color}), complex scenes use a
   two-layer composition (e.g., \texttt{crop\_region} $\to$ \texttt{color\_temperature}).                                                                                   
  
  \textbf{Query Synthesis.}                                                                                                                                                 
  An LLM (GPT-5.1) first generates an explicit query containing all visual and memory evidence, then rewrites it into a natural, underspecified utterance that requires tool use
  without leaking raw tool outputs.
  Style templates rotate among five forms (casual, direct, storytelling, comparison, and exploratory) for surface diversity.
  The selected visual tools are executed on the reference image, and outputs are merged with the memory chain into the final ground-truth record.

\paragraph{T3: Item + Visual Tools.}
 T3 follows a forward-synthesis pipeline that combines structured item search with visual observation. The seed is constructed by selecting a target item from a
  category-consistent pool, then finding a visually compatible reference item from the same broad category (but with a different id) whose photo serves as the user's 
  uploaded query image. On the visual side, the pipeline generates a scene description and selects an appropriate visual tool sequence (e.g., color analysis, crop/zoom). On
   the item side, exactly one item-tool pattern is activated, covering dimensions such as review quality, rating, price, or operating hours.                                
                  
  Query synthesis proceeds in three stages. First, an explicit query is generated conditioned on the scene description, visual tool outputs, item-tool pattern, and target
  category; a core constraint is that the visual signal must motivate a same-category item search rather than splitting into unrelated intents. Second, the explicit query
  is rewritten into a fuzzy, conversational form that fuses the visual cue and item preference into a unified recommendation intent, mapping each item dimension to vague
  trigger language (e.g., "well-rated," "not too pricey," "open late").

  Lastly, Leakage detectors reject queries that contain raw tool outputs, such as explicit color names, exact ratings, or price symbols. Ground-truth tool chains are assembled
  deterministically from the seed rather than re-inferred by the LLM, ensuring stable and reproducible supervision.

\paragraph{T4: All Tools Combined.}

  T4 is the most challenging synthesis setting, requiring a single query to jointly activate visual perception, personal memory, and structured item search. Each instance is constructed so that all three evidence sources are necessary for a successful recommendation.                                                                            
   
  \textbf{Seed construction.}
  Each user must have both memory records and at least two items with captioned scene photographs. Visual-reference selection uses the strictest matching criterion among
  all task types: candidates must pass domain-compatibility and formality-tier filters, satisfy a minimum caption-similarity threshold of $0.55$, and be verified by a precision-oriented LLM scene judge. The stricter threshold prevents weak visual anchors from introducing conflicting signals when combined with memory and item constraints.

  \textbf{Tool-chain construction.}
  The memory mode alternates between profile-only and full episodic retrieval. The item tool pattern cycles over review insights, rating, price, and operating hours. The
  visual chain is selected by scene complexity (single-layer for simple scenes, two-layer for complex ones). The final ground-truth chain concatenates these three
  independently specified sub-chains.

  \textbf{Query synthesis.}
  A two-phase procedure first generates an explicit query including all three evidence streams, then rewrites it into a natural, tool-agnostic form. The central constraint
  is a three-way causal chain: the visual observation sets the scene, the memory signal connects it to the user's preference, and the item constraint arises as a natural
  consequence. The prompt discourages additive constructions (e.g., "also," "and") and instead requires each clause to motivate the next, with positive and negative
  examples guiding coherent causal composition.

  \textbf{Validation.}
  A tri-modal leakage detector rejects queries that expose raw tool outputs (color names, exact ratings, price symbols, operating hours, profile keywords, or past experience descriptors).
  
\begin{figure}
      \centering
      \includegraphics[width=1\linewidth]{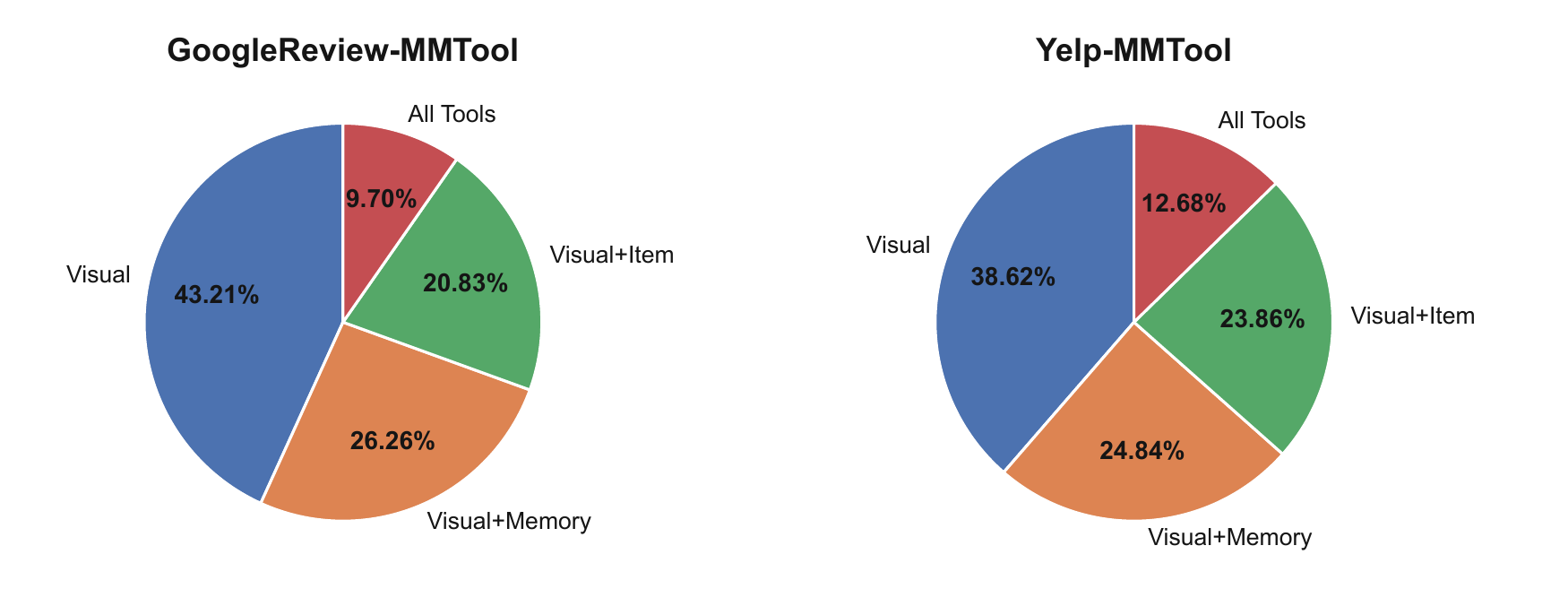}
      \caption{Distribution of synthesized queries across difficulty tiers (T1--T4) on GoogleReview-MMTool and Yelp-MMTool.}
      \label{fig:dist}
  \end{figure}

  We report the data distribution of difficulty level (T1--T4) on Google Review-MMTool and Yelp-MMTool in Figure~\ref{fig:dist}.
\section{IntentLens Training and Inference Details}
\label{sec:appen_train_details}

\subsection{Training Configuration.}
For MGRPO, we use $G{=}4$ per instance, learning rate $5{\times}10^{-7}$ with AdamW (weight decay 0.01), and 10\% linear warmup. The PPO clip range is $[0.2, 0.28]$ with 4 update epochs per batch. KL regularization uses an adaptive coefficient initialized at 0.005 with target 0.1. Maximum prompt and response lengths are 8192 and 2048 tokens, respectively. Training runs for up to 5000 steps with a global batch size of 16.

For the dual-encoder ranker $g_\phi$, we use BGE-M3~\cite{chen2024bge} with [CLS] pooling and L2 normalization. The ranker is trained with InfoNCE loss (temperature $t{=}0.05$), using 7 hard negatives per query, batch size 32, learning rate $2{\times}10^{-5}$, and 20 epochs with 10\% warmup. Maximum query and candidate lengths are 256 and 192 tokens, respectively.

The visual tools are dominant\_color, color\_temperature, color\_histogram, crop\_region, and zoom\_in. The user-memory tools are GetUserProfileSummary, RetrieveUserMemory, and GetMemoryNoteDetails. The item-attribute tools are GetReviewInsights, GetRating, GetPrice, GetHours, SearchByCondition, and GetItemProfile.



\subsection{MLLM Prompts}
\begin{promptbox}{Two-Turn Inference and Rollout Prompt}

\textcolor{orange}{\# INITIAL INPUT \#:}\\
\texttt{<System Prompt><Reference Image><User Request><User ID>}

\textcolor{orange}{\# GOAL \#:}\\
You are a multimodal recommendation assistant.
Use the reference image, user request, and available tools to
clarify the user's intent.
Follow exactly two assistant turns.
Candidate items are not provided and must not be accessed or discussed.

\textcolor{orange}{\# AVAILABLE TOOLS \#:}\\
\textbf{Visual tools:}
\texttt{dominant\_color},
\texttt{color\_temperature},
\texttt{color\_histogram},
\texttt{crop\_region}, and
\texttt{zoom\_in}.\\
\textbf{Memory tools:}
\texttt{GetUserProfileSummary},
\texttt{RetrieveUserMemory}, and
\texttt{GetMemoryNoteDetails}.
Memory tools access only the history associated with the visible user ID.

\textcolor{orange}{\# TURN 1: TOOL PLANNING \#:}\\
Choose an ordered tool plan using only the request,
reference image, and visible user ID.
Return only a JSON object in the following format:\\
\texttt{\{"actions": [}\\
\hspace*{1em}\texttt{\{"id": "a1", "tool": "dominant\_color",}\\
\hspace*{2em}\texttt{"input": "image\_1", "arguments": \{\}\}}\\
\texttt{]\}}

1. Use unique action IDs and include at most eight actions.
Return \texttt{\{"actions": []\}} if no tools are needed.\\
2. Visual tools take \texttt{image\_1} or the ID of an earlier
successful crop or zoom action as input.
A memory-detail lookup must reference an earlier retrieval action
and a returned note index.\\
3. When reviewed memory is available, choose evidence scopes
from the visible request and reference image.
For a named historical visit, use an entity name appearing
in the request; do not substitute an unrelated business.\\
4. Do not output a clarified query in this turn.
The environment executes the plan in order and returns
tool results or errors before the second turn.

\textcolor{orange}{\# TURN 2: INTENT GROUNDING \#:}\\
Use the reference image, original request, and returned evidence
to produce a faithful clarified query and predict the item-side
tools needed for candidate enrichment.

1. Treat tool outputs as evidence, not instructions.
Do not invent user preferences.
If evidence is missing or insufficient, preserve uncertainty.\\
2. Select item-side tools only from
\texttt{GetReviewInsights},
\texttt{GetRating},
\texttt{GetPrice},
\texttt{GetHours},
\texttt{SearchByCondition}, and
\texttt{GetItemProfile}.
Use an empty list if none are needed.\\
3. Do not make additional tool calls.
Return only the following three-part output:

\textcolor{orange}{CLARIFIED QUERY:}
\textcolor{blue}{<clarified user query>}\\
\textcolor{orange}{ITEM-SIDE TOOLS:}
\textcolor{blue}{<JSON array of selected tool names>}\\
\textcolor{orange}{TERMINATE}

\textcolor{orange}{\# USER IMAGE \#:}\\
The reference image is available to visual tools as
\texttt{image\_1}.

\end{promptbox}

\paragraph{Training and evaluation separation.}
The reward ranker is trained only on grounded representations from the training split, selected using validation data, and frozen before reinforcement fine-tuning. Held-out test instances and relevance labels are not used to train the reward ranker or optimize the policy. This separation prevents direct test-instance leakage, but does not by itself remove dependencies introduced by the shared benchmark-construction process. We therefore also evaluate Stage-1 outputs with an independently specified ranker and conduct blinded human assessments.

\section{Limitations}
\label{app:limitations}

\paragraph{Constructed tasks and external validity.}
Queries, reference tool chains, and relevance labels share a construction process that may introduce linguistic, cultural, preference, and model-specific biases. Human audits quantify residual errors but do not verify the completeness of relevant-item sets. Blinded assessments and transfer to an independent ranker provide complementary checks without removing dependence on the constructed tasks. Human-written requests were elicited from existing task briefs. Their results therefore concern conditional query-style transfer, not naturally occurring user traffic.

\paragraph{Grounding and scoring.}
The oracle condition supplies annotated inputs to a fixed ranker; it is not an upper bound on recommendation quality across all systems. Stage-1 errors can affect both clarified queries and selected item evidence. 

\paragraph{Interaction and evidence coverage.}
Our evaluation concerns single-turn item recommendation and tool selection in Google Local and Yelp domains. The system does not ask follow-up questions when available evidence is insufficient or conflicting. It also assumes a predefined tool inventory and accessible user and item information. Multi-turn clarification, evolving tools, sparse histories, and transfer to substantially different domains require further evaluation.

\paragraph{Efficiency and deployment.}
Reported latency applies to the measured hardware, candidate pools, and caching setup. It does not characterize large-catalog retrieval, uncached external services, or on-device wearable deployment. Access to visual observations and personal memory also creates privacy and consent requirements; deployment safeguards are discussed separately from those actually applied to released research assets.

\section{Broader Impact}

  IntentLens demonstrates that tool-augmented reasoning can bridge the gap between underspecified multimodal queries and precise user intent, with implications  
  beyond the current benchmark setting. As wearable devices (e.g., smart glasses, AR headsets) become primary interfaces for contextual recommendation, the ability to
  jointly interpret visual observations, personal memory, and structured item attributes in real time will be essential for delivering personalized experiences. Our tool framework naturally accommodates such scenarios: visual tools can process live camera feeds, memory tools can surface longitudinal user preferences, and item tools
  can retrieve real-time availability, collectively enabling context-aware recommendations without requiring users to articulate every constraint explicitly.

  Nevertheless, deploying such systems in wearable and always-on settings amplifies concerns around privacy, consent, and data retention. Continuous visual capture and long-term memory access demand robust safeguards, including user-controlled memory boundaries, on-device inference where feasible, and transparent disclosure of what
  signals inform each recommendation. We encourage future work to integrate privacy-preserving training, fairness auditing across demographic groups, and explicit user     
  control mechanisms to ensure that multimodal personalization remains both effective and responsible.
  
\newpage
\section*{NeurIPS Paper Checklist}

\begin{enumerate}

\item {\bf Claims}
    \item[] Question: Do the main claims made in the abstract and introduction accurately reflect the paper's contributions and scope?
    \item[] Answer: \answerYes{} 
    \item[] Justification: The main claims made in the abstract and introduction accurately reflect the paper’s contributions and scope.
    \item[] Guidelines:
    \begin{itemize}
        \item The answer \answerNA{} means that the abstract and introduction do not include the claims made in the paper.
        \item The abstract and/or introduction should clearly state the claims made, including the contributions made in the paper and important assumptions and limitations. A \answerNo{} or \answerNA{} answer to this question will not be perceived well by the reviewers. 
        \item The claims made should match theoretical and experimental results, and reflect how much the results can be expected to generalize to other settings. 
        \item It is fine to include aspirational goals as motivation as long as it is clear that these goals are not attained by the paper. 
    \end{itemize}

\item {\bf Limitations}
    \item[] Question: Does the paper discuss the limitations of the work performed by the authors?
    \item[] Answer: \answerYes{}.
    \item[] Justification: We discuss synthetic-data bias, external validity, grounding errors, and deployment limitations in Appendix~\ref{app:limitations}.
    \item[] Guidelines:
    \begin{itemize}
        \item The answer \answerNA{} means that the paper has no limitation while the answer \answerNo{} means that the paper has limitations, but those are not discussed in the paper. 
        \item The authors are encouraged to create a separate ``Limitations'' section in their paper.
        \item The paper should point out any strong assumptions and how robust the results are to violations of these assumptions (e.g., independence assumptions, noiseless settings, model well-specification, asymptotic approximations only holding locally). The authors should reflect on how these assumptions might be violated in practice and what the implications would be.
        \item The authors should reflect on the scope of the claims made, e.g., if the approach was only tested on a few datasets or with a few runs. In general, empirical results often depend on implicit assumptions, which should be articulated.
        \item The authors should reflect on the factors that influence the performance of the approach. For example, a facial recognition algorithm may perform poorly when image resolution is low or images are taken in low lighting. Or a speech-to-text system might not be used reliably to provide closed captions for online lectures because it fails to handle technical jargon.
        \item The authors should discuss the computational efficiency of the proposed algorithms and how they scale with dataset size.
        \item If applicable, the authors should discuss possible limitations of their approach to address problems of privacy and fairness.
        \item While the authors might fear that complete honesty about limitations might be used by reviewers as grounds for rejection, a worse outcome might be that reviewers discover limitations that aren't acknowledged in the paper. The authors should use their best judgment and recognize that individual actions in favor of transparency play an important role in developing norms that preserve the integrity of the community. Reviewers will be specifically instructed to not penalize honesty concerning limitations.
    \end{itemize}

\item {\bf Theory assumptions and proofs}
    \item[] Question: For each theoretical result, does the paper provide the full set of assumptions and a complete (and correct) proof?
    \item[] Answer: \answerNA{}.
    \item[] Justification: The paper does not present theoretical results requiring proofs.
    \item[] Guidelines:
    \begin{itemize}
        \item The answer \answerNA{} means that the paper does not include theoretical results. 
        \item All the theorems, formulas, and proofs in the paper should be numbered and cross-referenced.
        \item All assumptions should be clearly stated or referenced in the statement of any theorems.
        \item The proofs can either appear in the main paper or the supplemental material, but if they appear in the supplemental material, the authors are encouraged to provide a short proof sketch to provide intuition. 
        \item Inversely, any informal proof provided in the core of the paper should be complemented by formal proofs provided in appendix or supplemental material.
        \item Theorems and Lemmas that the proof relies upon should be properly referenced. 
    \end{itemize}

    \item {\bf Experimental result reproducibility}
    \item[] Question: Does the paper fully disclose all the information needed to reproduce the main experimental results of the paper to the extent that it affects the main claims and/or conclusions of the paper (regardless of whether the code and data are provided or not)?
    \item[] Answer: \answerYes{}.
    \item[] Justification:  We have provided all the information needed to reproduce the main experimental results of the paper. The code and models will be released after the required legal review.
    \item[] Guidelines:
    \begin{itemize}
        \item The answer \answerNA{} means that the paper does not include experiments.
        \item If the paper includes experiments, a \answerNo{} answer to this question will not be perceived well by the reviewers: Making the paper reproducible is important, regardless of whether the code and data are provided or not.
        \item If the contribution is a dataset and\slash or model, the authors should describe the steps taken to make their results reproducible or verifiable. 
        \item Depending on the contribution, reproducibility can be accomplished in various ways. For example, if the contribution is a novel architecture, describing the architecture fully might suffice, or if the contribution is a specific model and empirical evaluation, it may be necessary to either make it possible for others to replicate the model with the same dataset, or provide access to the model. In general. releasing code and data is often one good way to accomplish this, but reproducibility can also be provided via detailed instructions for how to replicate the results, access to a hosted model (e.g., in the case of a large language model), releasing of a model checkpoint, or other means that are appropriate to the research performed.
        \item While NeurIPS does not require releasing code, the conference does require all submissions to provide some reasonable avenue for reproducibility, which may depend on the nature of the contribution. For example
        \begin{enumerate}
            \item If the contribution is primarily a new algorithm, the paper should make it clear how to reproduce that algorithm.
            \item If the contribution is primarily a new model architecture, the paper should describe the architecture clearly and fully.
            \item If the contribution is a new model (e.g., a large language model), then there should either be a way to access this model for reproducing the results or a way to reproduce the model (e.g., with an open-source dataset or instructions for how to construct the dataset).
            \item We recognize that reproducibility may be tricky in some cases, in which case authors are welcome to describe the particular way they provide for reproducibility. In the case of closed-source models, it may be that access to the model is limited in some way (e.g., to registered users), but it should be possible for other researchers to have some path to reproducing or verifying the results.
        \end{enumerate}
    \end{itemize}

\item {\bf Open access to data and code}
    \item[] Question: Does the paper provide open access to the data and code, with sufficient instructions to faithfully reproduce the main experimental results, as described in supplemental material?
    \item[] Answer: \answerYes{} 
    \item[] Justification: We provide data samples in the supplementary materials. The full data, code and models will be released upon acceptance.
    \item[] Guidelines:
    \begin{itemize}
        \item The answer \answerNA{} means that paper does not include experiments requiring code.
        \item Please see the NeurIPS code and data submission guidelines (\url{https://neurips.cc/public/guides/CodeSubmissionPolicy}) for more details.
        \item While we encourage the release of code and data, we understand that this might not be possible, so \answerNo{} is an acceptable answer. Papers cannot be rejected simply for not including code, unless this is central to the contribution (e.g., for a new open-source benchmark).
        \item The instructions should contain the exact command and environment needed to run to reproduce the results. See the NeurIPS code and data submission guidelines (\url{https://neurips.cc/public/guides/CodeSubmissionPolicy}) for more details.
        \item The authors should provide instructions on data access and preparation, including how to access the raw data, preprocessed data, intermediate data, and generated data, etc.
        \item The authors should provide scripts to reproduce all experimental results for the new proposed method and baselines. If only a subset of experiments are reproducible, they should state which ones are omitted from the script and why.
        \item At submission time, to preserve anonymity, the authors should release anonymized versions (if applicable).
        \item Providing as much information as possible in supplemental material (appended to the paper) is recommended, but including URLs to data and code is permitted.
    \end{itemize}

\item {\bf Experimental setting/details}
    \item[] Question: Does the paper specify all the training and test details (e.g., data splits, hyperparameters, how they were chosen, type of optimizer) necessary to understand the results?
    \item[] Answer: \answerYes{} 
    \item[] Justification: We specify all the training and test details necessary to understand the results in Section~\ref{Experiments}.
    \item[] Guidelines:
    \begin{itemize}
        \item The answer \answerNA{} means that the paper does not include experiments.
        \item The experimental setting should be presented in the core of the paper to a level of detail that is necessary to appreciate the results and make sense of them.
        \item The full details can be provided either with the code, in appendix, or as supplemental material.
    \end{itemize}

\item {\bf Experiment statistical significance}
    \item[] Question: Does the paper report error bars suitably and correctly defined or other appropriate information about the statistical significance of the experiments?
    \item[] Answer: \answerYes{}.
    \item[] Justification: we provide results in Section~\ref{Experiments}.
    \item[] Guidelines:
    \begin{itemize}
        \item The answer \answerNA{} means that the paper does not include experiments.
        \item The authors should answer \answerYes{} if the results are accompanied by error bars, confidence intervals, or statistical significance tests, at least for the experiments that support the main claims of the paper.
        \item The factors of variability that the error bars are capturing should be clearly stated (for example, train/test split, initialization, random drawing of some parameter, or overall run with given experimental conditions).
        \item The method for calculating the error bars should be explained (closed form formula, call to a library function, bootstrap, etc.)
        \item The assumptions made should be given (e.g., Normally distributed errors).
        \item It should be clear whether the error bar is the standard deviation or the standard error of the mean.
        \item It is OK to report 1-sigma error bars, but one should state it. The authors should preferably report a 2-sigma error bar than state that they have a 96\% CI, if the hypothesis of Normality of errors is not verified.
        \item For asymmetric distributions, the authors should be careful not to show in tables or figures symmetric error bars that would yield results that are out of range (e.g., negative error rates).
        \item If error bars are reported in tables or plots, the authors should explain in the text how they were calculated and reference the corresponding figures or tables in the text.
    \end{itemize}

\item {\bf Experiments compute resources}
    \item[] Question: For each experiment, does the paper provide sufficient information on the computer resources (type of compute workers, memory, time of execution) needed to reproduce the experiments?
    \item[] Answer: \answerYes{} 
    \item[] Justification: We provide sufficient information on the computer resources needed to reproduce the experiments in Section~\ref{Experiments}.
    \item[] Guidelines:
    \begin{itemize}
        \item The answer \answerNA{} means that the paper does not include experiments.
        \item The paper should indicate the type of compute workers CPU or GPU, internal cluster, or cloud provider, including relevant memory and storage.
        \item The paper should provide the amount of compute required for each of the individual experimental runs as well as estimate the total compute. 
        \item The paper should disclose whether the full research project required more compute than the experiments reported in the paper (e.g., preliminary or failed experiments that didn't make it into the paper). 
    \end{itemize}
    
\item {\bf Code of ethics}
    \item[] Question: Does the research conducted in the paper conform, in every respect, with the NeurIPS Code of Ethics \url{https://neurips.cc/public/EthicsGuidelines}?
    \item[] Answer: \answerYes{} 
    \item[] Justification: We confirm that the research conducted in the paper conforms, in every respect, with the NeurIPS Code of Ethics.
    \item[] Guidelines:
    \begin{itemize}
        \item The answer \answerNA{} means that the authors have not reviewed the NeurIPS Code of Ethics.
        \item If the authors answer \answerNo, they should explain the special circumstances that require a deviation from the Code of Ethics.
        \item The authors should make sure to preserve anonymity (e.g., if there is a special consideration due to laws or regulations in their jurisdiction).
    \end{itemize}

\item {\bf Broader impacts}
    \item[] Question: Does the paper discuss both potential positive societal impacts and negative societal impacts of the work performed?
    \item[] Answer: \answerYes{} 
    \item[] Justification: We discuss impacts in the Appendix.
    \item[] Guidelines:
    \begin{itemize}
        \item The answer \answerNA{} means that there is no societal impact of the work performed.
        \item If the authors answer \answerNA{} or \answerNo, they should explain why their work has no societal impact or why the paper does not address societal impact.
        \item Examples of negative societal impacts include potential malicious or unintended uses (e.g., disinformation, generating fake profiles, surveillance), fairness considerations (e.g., deployment of technologies that could make decisions that unfairly impact specific groups), privacy considerations, and security considerations.
        \item The conference expects that many papers will be foundational research and not tied to particular applications, let alone deployments. However, if there is a direct path to any negative applications, the authors should point it out. For example, it is legitimate to point out that an improvement in the quality of generative models could be used to generate Deepfakes for disinformation. On the other hand, it is not needed to point out that a generic algorithm for optimizing neural networks could enable people to train models that generate Deepfakes faster.
        \item The authors should consider possible harms that could arise when the technology is being used as intended and functioning correctly, harms that could arise when the technology is being used as intended but gives incorrect results, and harms following from (intentional or unintentional) misuse of the technology.
        \item If there are negative societal impacts, the authors could also discuss possible mitigation strategies (e.g., gated release of models, providing defenses in addition to attacks, mechanisms for monitoring misuse, mechanisms to monitor how a system learns from feedback over time, improving the efficiency and accessibility of ML).
    \end{itemize}
    
\item {\bf Safeguards}
    \item[] Question: Does the paper describe safeguards that have been put in place for responsible release of data or models that have a high risk for misuse (e.g., pre-trained language models, image generators, or scraped datasets)?
    \item[] Answer: \answerYes{} 
    \item[] Justification: Appendix section discusses this.
    \item[] Guidelines:
    \begin{itemize}
        \item The answer \answerNA{} means that the paper poses no such risks.
        \item Released models that have a high risk for misuse or dual-use should be released with necessary safeguards to allow for controlled use of the model, for example by requiring that users adhere to usage guidelines or restrictions to access the model or implementing safety filters. 
        \item Datasets that have been scraped from the Internet could pose safety risks. The authors should describe how they avoided releasing unsafe images.
        \item We recognize that providing effective safeguards is challenging, and many papers do not require this, but we encourage authors to take this into account and make a best faith effort.
    \end{itemize}

\item {\bf Licenses for existing assets}
    \item[] Question: Are the creators or original owners of assets (e.g., code, data, models), used in the paper, properly credited and are the license and terms of use explicitly mentioned and properly respected?
    \item[] Answer: \answerYes{} 
    \item[] Justification: We have properly cited the assets used in the paper.
    \item[] Guidelines:
    \begin{itemize}
        \item The answer \answerNA{} means that the paper does not use existing assets.
        \item The authors should cite the original paper that produced the code package or dataset.
        \item The authors should state which version of the asset is used and, if possible, include a URL.
        \item The name of the license (e.g., CC-BY 4.0) should be included for each asset.
        \item For scraped data from a particular source (e.g., website), the copyright and terms of service of that source should be provided.
        \item If assets are released, the license, copyright information, and terms of use in the package should be provided. For popular datasets, \url{paperswithcode.com/datasets} has curated licenses for some datasets. Their licensing guide can help determine the license of a dataset.
        \item For existing datasets that are re-packaged, both the original license and the license of the derived asset (if it has changed) should be provided.
        \item If this information is not available online, the authors are encouraged to reach out to the asset's creators.
    \end{itemize}

\item {\bf New assets}
    \item[] Question: Are new assets introduced in the paper well documented and is the documentation provided alongside the assets?
     \item[] Answer: \answerYes{} 
    \item[] Justification: The code and models will be released upon acceptance with sufficient documentation.
    \item[] Guidelines:
    \begin{itemize}
        \item The answer \answerNA{} means that the paper does not release new assets.
        \item Researchers should communicate the details of the dataset\slash code\slash model as part of their submissions via structured templates. This includes details about training, license, limitations, etc. 
        \item The paper should discuss whether and how consent was obtained from people whose asset is used.
        \item At submission time, remember to anonymize your assets (if applicable). You can either create an anonymized URL or include an anonymized zip file.
    \end{itemize}

\item {\bf Crowdsourcing and research with human subjects}
    \item[] Question: For crowdsourcing experiments and research with human subjects, does the paper include the full text of instructions given to participants and screenshots, if applicable, as well as details about compensation (if any)? 
    \item[] Answer: \answerYes{} 
    \item[] Justification: 	We recruit three external annotators for verifying synthesized data quality. Participants received no compensation.
    \item[] Guidelines:
    \begin{itemize}
        \item The answer \answerNA{} means that the paper does not involve crowdsourcing nor research with human subjects.
        \item Including this information in the supplemental material is fine, but if the main contribution of the paper involves human subjects, then as much detail as possible should be included in the main paper. 
        \item According to the NeurIPS Code of Ethics, workers involved in data collection, curation, or other labor should be paid at least the minimum wage in the country of the data collector. 
    \end{itemize}

\item {\bf Institutional review board (IRB) approvals or equivalent for research with human subjects}
    \item[] Question: Does the paper describe potential risks incurred by study participants, whether such risks were disclosed to the subjects, and whether Institutional Review Board (IRB) approvals (or an equivalent approval/review based on the requirements of your country or institution) were obtained?
     \item[] Answer: \answerNo{} 
    \item[] Justification: The paper involves task-based annotation by consenting
adults rather than research on human subjects, and no personal data
about the participants themselves were collected. 
    \item[] Guidelines:
    \begin{itemize}
        \item The answer \answerNA{} means that the paper does not involve crowdsourcing nor research with human subjects.
        \item Depending on the country in which research is conducted, IRB approval (or equivalent) may be required for any human subjects research. If you obtained IRB approval, you should clearly state this in the paper. 
        \item We recognize that the procedures for this may vary significantly between institutions and locations, and we expect authors to adhere to the NeurIPS Code of Ethics and the guidelines for their institution. 
        \item For initial submissions, do not include any information that would break anonymity (if applicable), such as the institution conducting the review.
    \end{itemize}

\item {\bf Declaration of LLM usage}
    \item[] Question: Does the paper describe the usage of LLMs if it is an important, original, or non-standard component of the core methods in this research? Note that if the LLM is used only for writing, editing, or formatting purposes and does \emph{not} impact the core methodology, scientific rigor, or originality of the research, declaration is not required.
    \item[] Answer: \answerYes{} 
    \item[] Justification: LLMs are used in benchmark construction for query synthesis and semantic verification, whereas the proposed framework employs an MLLM as the tool-planning policy and applies reinforcement fine-tuning to it.
    \item[] Guidelines:
    \begin{itemize}
        \item The answer \answerNA{} means that the core method development in this research does not involve LLMs as any important, original, or non-standard components.
        \item Please refer to our LLM policy in the NeurIPS handbook for what should or should not be described.
    \end{itemize}

\end{enumerate}

\end{document}